\documentclass[11pt, a4paper]{article}

\usepackage{jheppub}
\usepackage{amsmath, amsfonts, amsthm, amssymb, graphicx}
\usepackage[normalem]{ulem}
\usepackage{color}
\usepackage{cancel}
\usepackage{float}
\usepackage[utf8]{inputenc}
\usepackage{lipsum}
\usepackage{centernot}
\usepackage{ulem}
\usepackage{comment}
\usepackage{mathtools}
\usepackage{enumitem}
\usepackage{feynmp}
\usepackage{comment}
\usepackage[sort&compress]{natbib}

\newcommand{\x}{\relax\ifmmode \mathcal{X} \else $\mathcal{X}$ \fi}

\def\be{\begin{equation}}
\def\ee{\end{equation}}
\def\ben{\begin{equation*}}
\def\een{\end{equation*}}

\def\del{\partial}

\newcommand{\vol}{\relax\ifmmode \mathcal{V} \else $\mathcal{V}$ \fi}

\newcommand{\vmin}{\relax\ifmmode V^{(n)}_{\text{min}}\else $V^{(n)}_{\text{min}}$ \fi}
\def\hess#1{{\relax\ifmmode {\mathbf H}_{#1} \else ${\mathbf H}_{#1}$ \fi}}

\title{From inflation to quintessence: \\ a history of the universe in string theory}

\author[a,b]{Michele Cicoli,} 
\author[c,d]{Francesc Cunillera,} 
\author[c,d]{Antonio Padilla,} 
\author[a,b]{Francisco G. Pedro} 

\affiliation[a]{Dipartimento di Fisica e Astronomia, Università di Bologna, via Irnerio 46, 40126 Bologna, Italy} 
\affiliation[b]{INFN, Sezione di Bologna, viale Berti Pichat 6/2, 40127 Bologna, Italy} 
\affiliation[c]{School of Physics and Astronomy, University of Nottingham, Nottingham NG7 2RD, UK} 
\affiliation[d]{Nottingham Centre of Gravity, University of Nottingham, Nottingham NG7 2RD, UK}

\emailAdd{michele.cicoli@unibo.it}
\emailAdd{francesc.cunilleragarcia@nottingham.ac.uk}
\emailAdd{antonio.padilla@nottingham.ac.uk}
\emailAdd{francisco.soares@unibo.it}

\abstract{We present a type IIB 4D string model with stabilised moduli which is able to describe the history of the universe from inflation to quintessence. The underlying Calabi-Yau volume is controlled by two moduli which are stabilised by perturbative effects. The lighter of them drives Fibre Inflation at a large energy scale. The two associated axions are ultra-light since they are lifted only at the non-perturbative level. The lighter of them can drive quintessence if its decay constant is large enough to prevent quantum diffusion during inflation from ruining the initial conditions. The right dark energy scale can be obtained via a large suppression from poly-instanton effects. The heavier axion gives a negligible contribution to dark matter since it starts oscillating after matter-radiation equality. If instead none of the two axions has a large decay constant, a mild alignment allows the lighter axion to drive quintessence, while the heavier can be at most a few percent of dark matter due to isocurvature and UV bounds. In both cases dark matter can also come from either primordial black holes or the QCD axion.}

\begin{document}

\maketitle

\section{Introduction}

A slew of cosmological observations, from the temperature anisotropies in the cosmic microwave background radiation \cite{CMB} to the light of distant supernovae \cite{Sn1, Sn2}, point to a cosmic history bookended by two periods of accelerated expansion. At early times we had inflation, answering cosmological puzzles such as the horizon problem and providing the seed for cosmic structure \cite{daniel}. At late times, we have dark energy, driven by either vacuum energy or a quintessence field in slow-roll \cite{ed, quin}. There is a huge hierarchy of scales between the two epochs presenting challenges for model builders, with the low scale of dark energy raising further questions about naturalness \cite{cliff,tony}, whether it is driven by vacuum energy or quintessence. In between these two epochs, the universe evolved through a period of radiation domination, followed by matter, the latter composed mostly of dark matter whose microscopic origin has yet to be established \cite{DM1, DM2}. 

Identifying a consistent cosmological model inline with this cosmic history is one of the most important goals of string cosmology (see \cite{Cicoli:2023opf} for a recent review). The greatest challenge to this comes from realising accelerated expansion during inflation and today, together with the right phenomenological scales. The main approach is to focus on the low-energy limit of Calabi-Yau (CY) compactifications where the underlying supersymmetry, large volume and weak string coupling guarantee control over the effective field theory. This framework has led to several proposals for obtaining accelerated expansion from a de Sitter (dS) vacuum, including anti-branes in warped throats \cite{kklt, Aparicio:2015psl, Crino:2020qwk}, D-terms \cite{Burgess:2003ic, Braun:2015pza}, T-branes \cite{Cicoli:2015ylx, Cicoli:2012vw, Cicoli:2013mpa, Cicoli:2013cha, Cicoli:2017shd, Cicoli:2021dhg}, F-terms of the complex structure moduli \cite{Gallego:2017dvd}, non-perturbative effects at singularities \cite{Cicoli:2012fh}, and $\alpha'$ contributions \cite{Westphal:2006tn}. Nonetheless, the existence of stable de Sitter vacua in controlled string compactifications has been challenged via some conjectures \cite{nodS1, nodS2,nodS3} and no-go theorems in certain approximations \cite{Maldacena:2000mw,Green:2011cn, Gautason:2012tb,Kutasov:2015eba,Quigley:2015jia,Dine:2020vmr,Montero:2020rpl,Cunillera:2021fbc}. For further discussion on the status of de Sitter vacua in string theory, see \cite{vrdS} and \cite{Cicoli:2018kdo}. 

This does not mean that it is any easier to obtain a dynamical model of accelerated expansion since in \cite{Cicoli:2021fsd,Cicoli:2021skd} we showed that quintessence models have the same control issues as dS constructions with however extra phenomenological constraints. In fact, building on earlier work of a similar spirit \cite{Hertzberg:2007wc,Garg:2018zdg,Ibe:2018ffn,ValeixoBento:2020ujr, HebQ}, in \cite {Cicoli:2021fsd} we showed that accelerated expansion is not possible in the parametrically controlled regime at the boundary of moduli space.\footnote{In \cite{Calderon-Infante:2022nxb} the same problem is addressed in F-theory. While the approach certainly has its merits, the authors merely look for acceleration rather than observationally viable acceleration, a much more constraining requirement. Furthermore, as the authors acknowledge, \cite{Calderon-Infante:2022nxb} does not tackle moduli stabilisation, without which robust observational signatures cannot be derived from the underlying string model.} This negative result has been extended to the multi-field case with canonical kinetic terms in \cite{Shiu:2023nph, Shiu:2023fhb}. Steep potentials could, however, still provide accelerating solutions once the kinetic coupling to axionic degrees of freedom is taken into account \cite{Cicoli:2020cfj,Cicoli:2020noz,Shiu:2024sbe}, even if no solution consistent with late time observations has been found so far \cite{Brinkmann:2022oxy}.

This analysis shows that quintessence needs to be realised in the bulk of moduli space where numerical control could still be retained in a consistent expansion in small parameters (e.g. inverse volume, string coupling) \cite{Cicoli:2021skd}. However, even if late time acceleration might naively seem relatively easy to realise since it requires only one efolding of acceleration, as opposed to 50 to 60 efoldings at early times, the fact that the quintessence field has to be ultra-light with mass $m\sim H_0\sim 10^{-60}\,M_p$ yields a few model building challenges \cite{Hebecker:2019csg}: ($i$) how can fifth forces be avoided? ($ii$) how can such a low mass scale be radiatively stable? ($iii$) how can this tiny mass be obtained keeping the string scale and the soft terms above the TeV scale?

In \cite{Cicoli:2021skd} we argued that these challenges can be successfully addressed if the dark energy field is an axion which enjoys a perturbatively exact shift symmetry. Being an axion, this field would not lead to any observable fifth force. Moreover, the smallness of its mass with respect to the one of the other moduli (which need to be heavier than about $1$ meV to avoid fifth forces) and its radiative stability are naturally explained by the fact that axions acquire mass only via exponentially suppressed non-perturbative effects. Due to this observation, in \cite{Cicoli:2021skd} we developed a blueprint for a consistent model of quintessence in perturbative string theory which can be presented generically in terms of  the underlying scalar potential
\begin{equation}
\label{breakdown}
V=V_\text{vol}(\mathcal{V})+V_\text{inf}(\sigma, \mathcal{V})+V_\text{late}(\phi, \mathcal{V})
\end{equation}
This potential has a hierarchical structure since it is split into a leading order contribution for the volume mode $V_\text{vol}(\mathcal{V})$, a correction for inflation at early times $V_\text{inf}(\sigma, \mathcal{V})$, and a much smaller potential for quintessence at late times, $V_\text{late}(\phi, \mathcal{V})$.  Here the volume of the internal Calabi-Yau manifold is given by $\mathcal{V}$, the inflaton by $\sigma$ and the quintessence field by $\phi$.

As we cannot achieve phenomenologically viable acceleration at the boundary of moduli space, we step into the bulk and expand in small parameters. In particular, the blueprint states that:
\begin{itemize}
\item at leading order, $V_\text{vol}(\mathcal{V})$ should admit a non-supersymmetric near Minkowski vacuum with two flat directions, $\sigma$ and $\phi$.

\item at sub-leading order, $V_\text{inf}(\sigma, \mathcal{V})$ should contain an inflationary plateau at high energies $V_\text{inf}(\sigma, \mathcal{V})\simeq 3 H_{\rm inf}^2 M_p^2\gtrsim (\text{MeV})^4$, while $\phi$ should still remain flat. 

\item at sub-sub-leading order, $V_\text{late}(\phi, \mathcal{V})$ should be generated by non-perturbative corrections that lift $\phi$ at the cosmological constant scale, $V_\text{late}(\phi, \mathcal{V})\simeq 3 H_0^2 M_p^2\simeq (\text{meV})^4$. 
\end{itemize}
Note that $V_\text{inf}(\sigma, \mathcal{V})\gg V_\text{late}(\phi, \mathcal{V})$ can be naturally realised if the inflationary potential is generated by perturbative effects, as in \cite{Cicoli:2008gp, Broy:2015zba, Burgess:2016owb, Cicoli:2016chb,Cicoli:2016xae,Cicoli:2017axo,Bera:2024sbx,Bansal:2024uzr}. In this perspective, the hierarchy between the inflationary and the dark energy scales would be induced by the underlying hierarchy between perturbative and non-perturbative corrections to the effective action. Moreover, $V_{\rm vol}(\mathcal{V})\gg V_\text{inf}(\sigma, \mathcal{V})$ guarantees that these models are not at risk from decompactification when inflationary corrections are included \cite{KL}. Note, however, that the inflaton does not necessarily need  to be distinct from the volume mode, as in \cite{Linde:2007jn,Conlon:2008cj,Cicoli:2015wja, Antoniadis:2020stf,Antoniadis:2021lhi,Bera:2024zsk}. In this case, the condition $V_{\rm vol}(\mathcal{V})\gg V_\text{inf}(\sigma, \mathcal{V})$ can clearly be relaxed.

The purpose of this paper is to present an explicit model inspired by our blueprint in case future observations favour dynamical dark energy over dS models, with the recent results of the DESI collaboration suggesting this could well be the case \cite{DESI:2024mwx}.   We work with type IIB string theory and focus on the Large Volume Scenario (LVS) on a K3-fibred Calabi-Yau \cite{Balasubramanian:2005zx,Cicoli:2008va,Cicoli:2011it}. The volume is stabilised at leading order in a non-supersymmetric Minkowski vacuum. Thanks to the fibration structure, the large K\"ahler moduli admit a direction that is flat even in the presence of $\alpha'^3$ corrections. This flat direction can then be lifted using loop corrections to the K\"ahler potential or higher derivative contributions to the scalar potential, giving rise to Fibre Inflation \cite{Cicoli:2008gp, Broy:2015zba, Burgess:2016owb, Cicoli:2016chb, Cicoli:2016xae, Cicoli:2017axo, Bera:2024sbx} at around $10^{13}$ GeV. As studied in \cite{Cicoli:2018cgu,Cicoli:2023njy,Cicoli:2022uqa}, the perturbative decay of the inflaton after the end of inflation reheats the Standard Model which can be realised with either D7 or D3-branes. The post-inflationary evolution determines the number of efoldings of inflation which turns out to be $N_e\simeq 52$. A small amount of dark radiation compatible with data is sourced by relativistic ultra-light axions produced from the inflaton decay. Dark matter can come from primordial black holes \cite{Cicoli:2018asa, Cicoli:2022sih}, or a QCD axion on D3-branes with a decay constant below $H_{\rm inf}$, avoiding isocurvature bounds \cite{Cicoli:2022uqa}.  

The dynamics of the late universe is generated from non-perturbative corrections to the superpotential in order to have the required  hierarchy between inflation and dark energy, making the quintessence field as light as the Hubble scale today. For the fibred case considered here, this generates a potential for the two axions associated to the two large moduli. Generically, the potential for an axion is too steep to be able to drive quintessence even if, in principle, one could still obtain an accelerating solution if the axion sits very close to the maximum of its potential. As shown in \cite{Kaloper:2005aj, Cicoli:2021skd}, quantum diffusion during inflation would very quickly move the axion away from the maximum. 

In order to avoid this diffusion problem, for $H_{\rm inf}\gtrsim 10^{13}$ GeV, one needs a relatively large axion decay constant, $f\gtrsim 0.1\,M_p$, which can still be achieved in a controlled effective field theory where all moduli are large in string units. However, for a standard instanton with action $S$, the weak gravity conjecture gives $f\, S\lesssim \mathcal{O}(1)$ \cite{Arkani-Hamed:2006emk}, implying that $S$ would be too small to reproduce the correct cosmological constant scale. We overcome this problem by exploiting poly-instanton corrections to the superpotential \cite{Blumenhagen:2008ji,Blumenhagen:2012kz,Blumenhagen:2012ue,Lust:2013kt} which can lead to a very suppressed axion potential even for a large decay constant.\footnote{Poly-instanton corrections have already been used in \cite{Cicoli:2012tz} to build a quintessence model where however the dark energy field is a saxion.} These corrections naturally generate a hierarchy between the two axions. The lighter of them drives quintessence at the right dark energy scale, while the heavier is a spectator field with negligible contribution to dark matter since it starts oscillating after matter-radiation equality. 

We also analyse the case where the values of the UV parameters are such that the decay constants of both axions are of order $0.01\,M_p$. In this case, applying to quintessence the axion alignment mechanism developed for inflation in \cite{hep-ph/0409138,Long:2014dta,2106.09853}, with a modest amount of alignment we can make the decay constant of one of the two axions large enough to sustain acceleration for a large range of initial conditions that are not destroyed by quantum diffusion during inflation. The lighter axion plays the role of quintessence while its heavier cousin corresponds to a small fraction of dark matter (around at most $5\%$). Concerns about violations of the weak gravity conjecture are avoided using an additional instanton correction that only has a negligible effect on the dark energy dynamics \cite{Rudelius:2015xta, 2106.09853}. 

The rest of this paper is organised as follows: in Sec. \ref{sec:setup} we present our main setup where the axion dynamics is determined by poly-instantons. In Sec. \ref{sec:constraints} we discuss the relevant phenomenological constraints on the parameters of our model. In particular, this includes constraints on inflation, reheating, dark energy and dark matter. Finally, in Sec. \ref{sec:conc}, we conclude. App. \ref{AppAlignment} provides the details of the alternative scenario based on axion alignment.

\section{The setup}
\label{sec:setup}

We consider the 4D low-energy supergravity effective limit of type IIB Calabi-Yau compactifications. As the complex structure moduli and the dilaton are stabilised by fluxes at tree-level, we focus just on the K\"ahler moduli $T_i=\tau_i+i \theta_i$. Their Lagrangian is given by
\be 
\label{L}
\mathcal{L}=-K_{i \bar j} \del_\mu T_i \del^\mu \bar T_{\bar j}-V\, .
\ee
The K\"ahler metric $K_{i \bar j}=\partial_i \partial_{\bar j} K$ is given in terms of the K\"ahler potential $K$, with $\partial_i$ denoting partial differentiation with respect to the K\"ahler modulus $T_i$. The standard F-term scalar potential is given in terms $K$ and the superpotential $W$ as
\be
V=e^K\left[ K^{i \bar j} D_i W D_{\bar j} \bar W-3 |W|^2\right]\, ,
\ee 
where $D_i W\equiv\partial_iW +W\partial_i K$ is the K\"ahler covariant derivative and $K^{i \bar j} $ is the inverse of the K\"ahler metric.  

Our proposed model is an LVS compactification on a fibred Calabi-Yau. As usual, we work with the dimensionless internal volume in Einstein frame, expressed in units of the string length $\ell_s=2 \pi \sqrt{\alpha'}$. In terms of $2$-cycles, this is given by (see  \cite{Cicoli:2011it,Cicoli:2016xae,Cicoli:2017axo,Bera:2024sbx} for explicit K3-fibred CY examples):
\be
\vol=\frac{k}{2}\,t_1 t_2^2+\frac{\hat{k}}{6}\,t_s^3\,,
\label{VolForm}
\ee
where $t_s<0$, and $k$ and $\hat{k}$ are the CY intersection numbers which are positive integers. We can also express the volume in terms of the corresponding $4$-cycles, by writing
\be
\vol =\frac{1}{\sqrt{2k}}\, \sqrt{\tau_1} \tau_2 - \frac13\sqrt{\frac{2}{\hat{k}}}\, \tau_s^{3/2}\,,
\label{CYvol}
\ee
and identifying
\begin{equation}
\tau_1=\frac{\del \vol}{\del t_1}=\frac{k}{2}\,t_2^2\,, \qquad \tau_2=\frac{\del \vol}{\del t_2}=k\, t_1 t_2\,, \qquad \tau_s=\frac{\del \vol}{\del t_s}=\frac{\hat{k}}{2}\,t_s^2\,.
\end{equation}
As shown in \cite{oguiso}, a CY threefold whose volume (\ref{VolForm}) is linear in $t_1$ features a $\tau_1$ $4$-cycle which is a K3 or $\mathbb{T}^4$ divisor fibred over a $\mathbb{P}^1$ base with volume given by $t_1$.

At tree-level, the K\"ahler potential is given by $K=-2\ln \vol$ and the superpotential by $W=W_0$. This yields a vanishing scalar potential due to the well known no-scale structure. We therefore include a number of corrections to lift the K\"ahler moduli, specifically:
\begin{itemize}
\item Perturbative corrections to the K\"ahler potential:
\be
K \to -2\ln \left(\vol+\frac{\xi}{2 g_s^{3/2}} \right) +K_{g_s}
\ee
coming from $\alpha'^3$ terms in the effective action as well as string loops. The correction proportional to $\xi$ comes from the dimensional reduction of 10D $\alpha'^3$ terms which scale schematically as $R^4 + R^3 G_3^2$. It is controlled by $\xi=-\frac{\zeta(3)\chi}{2 (2\pi)^3}$, where $\chi$ is the Euler number of the Calabi-Yau \cite{Becker:2002nn}.\footnote{Note that $N=1$ $\alpha'^3$ corrections can cause a shift of the CY Euler number \cite{Minasian:2015bxa}.} The behaviour of the loop corrections has been conjectured from insights from a toroidal computation and effective field theory arguments, leading to \cite{Berg:2005ja,vonGersdorff:2005bf,Berg:2007wt,Cicoli:2007xp,Gao:2022uop}:
\be
K_{g_s}=\sum_i g_s \frac{C_i(U,\bar{U})\, t_i^{\perp}}{\mathcal{V}}+ \sum_i \frac{\tilde{C}_i(U,\bar{U})}{t_i^{\cap}\, \mathcal{V}}\,.
\ee
In the closed string channel, these corrections can be seen to arise from the tree-level exchange of Kaluza-Klein strings between parallel D7-branes (with $t_i^\perp$ denoting the volume of the transverse $2$-cycle) and winding strings at the intersection among D7-branes (with $t_i^\cap$ denoting the  volume of the intersection locus). $C$ and $\tilde{C}$ are unknown functions of the complex structure moduli $U$, which can be considered as constants since the $U$-moduli are fixed at tree-level. For the fibred case at hand, the relevant $t_i^{\perp}$'s are $t_1$ and $t_2$, while $t_i^{\cap}$ corresponds just to $t_2$, the volume of the intersection between a D7-stack wrapping the $4$-cycle with volume $\tau_1$ and another D7-stack wrapping the divisor with volume $\tau_2$. 

\item Higher derivative corrections to the scalar potential $V\to V+ V_{\rm hd}$:
\be
V_{\rm hd}=- \frac{\lambda}{g_s^{3/2}} \frac{|W_0|^4}{\vol^4}\,\Pi_i t_i
\ee
coming from the dimensional reduction of 10D $\alpha'^3$ contributions which scale as $R^2 G_3^4$ \cite{Ciupke:2015msa}.\footnote{Throughout this paper we will always set the overall prefactor of the scalar potential equal to unity,  $e^{K_{\rm cs}}\,g_s/\left(8\pi\right)=1$.} Here $\lambda$ is an unknown combinatorial number that is expected to be negative and of order $10^{-4}$ in absolute value \cite{Cicoli:2023njy}, while the $\Pi_i$ are topological quantities with positive integer values. In particular, the $\tau_1$ modulus has $\Pi_1=24$ if the corresponding $4$-cycle is a K3 divisor, or $\Pi_1=0$ if it is a $\mathbb{T}^4$ \cite{Cicoli:2023njy}.

\item Non-perturbative corrections to the superpotential, $W \to W_0+ W_{\rm np}$:
\begin{eqnarray}
W_{\rm np}&=&A_s\,e^{-a_s T_s}+ A_2 \,e^{-a_2 T_2 \,+ A_1\,e^{-a_1 T_1}} \nonumber \\
&=& A_s\,e^{-a_s T_s}+ A_2 \,e^{-a_2 T_2}+ A_2 A_1 \,e^{-\left(a_2 T_2 + a_1 T_1\right)} + ...
\label{Wnp}
\end{eqnarray}
coming from Euclidean D3-brane instantons ($a_i=2\pi$) or gaugino condensation on a stack of $N_i$ D7-branes ($a_i=2\pi/N_i$) \cite{Blumenhagen:2009qh}. Note that the $T_1$-dependent non-perturbative correction is a poly-instanton contribution to $W$ that arises as an instanton correction to the gauge kinetic function of the D7-stack wrapping the $T_2$-cycle \cite{Blumenhagen:2008ji,Blumenhagen:2012kz,Blumenhagen:2012ue,Lust:2013kt}.\footnote{This D7-stack is fictitious when the $T_2$-cycle is wrapped by an instanton, i.e. when $N_2=1$.} Recall that the $\tau_1$ divisor can be either a K3 or a $\mathbb{T}^4$. However, as pointed out in \cite{Lust:2013kt}, only the K3 divisor has the right zero mode structure to generate a non-perturbative superpotential of the form (\ref{Wnp}). We shall therefore focus on this case and set $\Pi_1=24$ in what follows. In the second line of (\ref{Wnp}) we have neglected terms which are exponentially suppressed. Here we consider up to three stacks: one required to implement moduli stabilisation in LVS, wrapping only the small cycle modulus $T_s$; two more wrapping only the large cycle moduli, $T_1$ and $T_2$, giving rise to a rich axion phenomenology at late times.
\end{itemize}
When we bring all of this together, the kinetic part of the Lagrangian \eqref{L} takes the following form to leading order
\begin{equation}
\mathcal{L}_\text{kin} \simeq -\frac{1}{4\tau_1^2} \left[ (\del \tau_1)^2+(\del \theta_1)^2 \right]-\frac{1}{2\tau_2^2} \left[ (\del \tau_2)^2+(\del \theta_2)^2 \right]-\frac{1}{4\sqrt{2\hat{k}}\,\vol \sqrt{\tau_s}} \left[ (\del \tau_s)^2+(\del \theta_s)^2 \right].
\end{equation}
The scalar potential can also be computed at leading order, and then decomposed in the form along the lines of \eqref{breakdown}, with 
\be
V=V_\text{vol}+V_\text{inf}+V_\text{late}\,.
\ee
Here the leading order LVS piece required to stabilise the volume is given by
\be 
V_\text{vol}=\frac{\kappa}{\vol^n}+\frac{3 \xi |W_0|^2}{4 g_s^{3/2}\vol^3}-4 |W_0| A_s a_s\tau_s\, \frac{e^{-a_s \tau_s}}{\vol^2} \cos (a_s\theta_s)+4\sqrt{2 \hat{k}}\, A_s^2 a_s^2 \sqrt{\tau_s}\, \frac{e^{-2a_s\tau_s}}{\vol}\,,
\ee
where we have included an uplifting contribution with $0<n<3$ to achieve a Minkowski vacuum by an appropriate tuning of the positive parameter $\kappa$. In particular, just to quote some examples, $n=4/3$ for an anti-D3-brane at the tip of a warped throat \cite{kklt}, $n=8/3$ for T-branes \cite{Cicoli:2015ylx} and $n=2$ for complex structure F-terms \cite{Gallego:2017dvd}. Moreover, we have set the Planck mass $M_p=1$ and we have further assumed, without loss of generality, $W_0=-|W_0|$ and that $A_s$ is real and positive.

The inflationary correction is given by
\be 
\label{Vinf}
V_\text{inf}= \frac{|W_0|^2}{\vol^3} \left( \frac{ B_1\,|W_0|^2}{\tau_1}-\frac{\sqrt{2k}\,\tilde C}{\sqrt{\tau_1}}+\sqrt{\frac{2}{k}}\frac{B_2\,|W_0|^2\sqrt{\tau_1}}{\vol} \right),
\ee
where we have defined the constants $B_i\equiv |\lambda|\,\Pi_i/g_s^{3/2}$, for $i=1, 2$ (taking $\lambda<0$).  Note that Kaluza-Klein loop corrections enjoy an extended no-scale cancellation \cite{Cicoli:2007xp}, and so we did not include them. Taking these corrections into account would however not modify the inflationary picture qualitatively. In fact, KK loop corrections have been used to generate the inflationary potential in the original version of Fibre Inflation \cite{Cicoli:2008gp} and in \cite{Broy:2015zba,Cicoli:2016xae}.

Finally, the late time correction to the scalar potential is given by
\begin{eqnarray}
V_\text{late}= &-& \frac{4|W_0| A_2}{\vol^2}\,\left(a_2\,\tau_2 \right)\,e^{-a_2 \tau_2} \cos\left(a_2 \theta_2\right) \nonumber \\
&-& \frac{4|W_0| A_2 A_1}{\vol^2}\,\left(a_2\,\tau_2 +a_1 \tau_1\right)\,e^{-\left(a_2 \tau_2+a_1\tau_1\right)} \cos\left(a_2 \theta_2+a_1\theta_1\right)
\label{Vlate}
\end{eqnarray} 
where we have again assumed, without loss of generality, that $A_1$ and $A_2$ are both real and positive. Note that we have kept the leading order terms and dropped those that are relatively suppressed, at least in the limit of large volume and when we respect the hierarchies $\tau_2 \gg \tau_s$ and $\tau_1\gtrsim \tau_s$. 

We now consider the dynamics of this theory, including the stabilisation of the K\"ahler moduli and the dynamical rolling of some, giving accelerated expansion at both early and late times.

\subsection{Volume stabilisation}

We begin with the stabilisation of the volume.  To this end, we can neglect subleading contributions from $V_\text{inf}$ and $V_\text{late}$ and focus exclusively on $V_\text{vol}$.  The axion of the small $4$-cycle is fixed at $a_s\langle\theta_s\rangle=2c\pi$ with $c\in \mathbb{Z}$, with the corresponding square mass given by 
\be
m^2_{\theta_s}\simeq 8\sqrt{2\hat{k}}\, |W_0| A_s a_s^3 \tau_s^{3/2}\, \frac{e^{-a_s\tau_s}}{\vol}\, .
\ee
This leaves us with a leading order volume potential of the form
\be
V_\text{vol}=\frac{\kappa}{\vol^n}+\frac{3 \xi |W_0|^2}{4 g_s^{3/2} \vol^3}-4 |W_0| A_s a_s\tau_s \frac{e^{-a_s\tau_s}}{\vol^2} +4\sqrt{2\hat{k}}\, A_s^2 a_s^2\sqrt{\tau_s}\, \frac{e^{-2a_s\tau_s}}{\vol}\,.
\ee
We now minimise along the $\vol$ and $\tau_s$ directions, and tune the uplift $\kappa$ so that the potential vanishes at the minimum, $\del_\vol V_\text{vol}= \del_{\tau_s} V_\text{vol}=V_\text{vol}=0$. This is consistent with the assertion that the volume should be stabilised at a non-supersymmetric (near) Minkowski minimum, and yields three simultaneous equations that are solved by
\be 
\langle\tau_s\rangle \simeq \frac{\hat{k}^{1/3}(3\xi)^{2/3}}{2 g_s} \,, \quad \vol\simeq \frac{|W_0| \sqrt{\langle\tau_s\rangle} }{2\sqrt{2\hat{k}} A_s a_s}\,e^{a_s\langle\tau_s\rangle}\,, \quad \kappa\simeq \frac{9 \xi |W_0|^2}{8 (3-n)g_s^{3/2} a_s\langle\tau_s\rangle}\, \vol^{n-3}\,.
\label{sols}
\ee
After imposing this solution, the small axion mass goes as 
\be 
m_{\theta_s} \simeq \frac{|W_0| a_s\langle\tau_s\rangle}{\vol}\,,
\label{axionmass}
\ee
and so it is of order of the gravitino mass $m_{3/2} \simeq |W_0|/\vol$.
The masses for the saxions can be easily obtained after rewriting the kinetic Lagrangian in terms of the volume mode and the mode orthogonal to the volume, $\Sigma=\ln (\tau_1/\tau_2)/\sqrt{3} = \langle\Sigma\rangle+\sigma$, which we have expanded around its minimum $\langle\Sigma\rangle$. The kinetic terms for the saxions can now be written as
\be
\mathcal{L}_\text{kin} \supset -\frac12  \begin{pmatrix} \del \ln \vol, &\del \ln \tau_s \end{pmatrix} \mathcal{K} \begin{pmatrix} \del \ln \vol \\ \del \ln \tau_s \end{pmatrix} -\frac12 (\del \sigma)^2
\ee
where 
\be
\mathcal{K}=\begin{pmatrix} \frac23 & -\frac{\tau_s^{3/2}}{\sqrt{2\hat{k}}\, \vol} \\
-\frac{\tau_s^{3/2}}{\sqrt{2\hat{k}}\, \vol} & \frac{\tau_s^{3/2}}{2\sqrt{2\hat{k}}\, \vol}\end{pmatrix}\,.
\ee
The orthogonal mode, $\sigma$, is clearly massless and will only be lifted by inflationary corrections. To obtain the masses of the other two moduli, we compute the Hessian, $\mathcal{H}$, of $V_\text{vol}$ by differentiating with respect to $\ln \vol$ and $\ln \tau_s$, and compute the eigenvalues of the corresponding mass matrix $\mathcal{K}^{-1} \mathcal{H}$. The result for the large and small moduli is respectively:
\be 
m^2_\vol \simeq \sqrt{\frac{2}{\hat{k}}} \frac{\sqrt{\langle\tau_s\rangle} |W_0|^2}{3 a_s \vol^3} \qquad\text{and}\qquad m^2_{\tau_s} \simeq  \left(\frac{|W_0| a_s\langle\tau_s\rangle}{\vol}\right)^2\,.
\label{saxionmasses}
\ee

\subsection{The early universe} 
\label{sec:inflation}

We now consider the inflationary correction in the early universe. With $\tau_s$, $\theta_s$ and the volume already stabilised, we take the inflaton to be the  mode orthogonal to the volume, namely $\Sigma=\ln (\tau_1/\tau_2)/\sqrt{3} = \langle\Sigma\rangle+\sigma$. The inflaton potential can now be written as 
\be 
V_\text{inf}(\sigma)= \frac{B_1|W_0|^4}{(2k)^{1/3}\,\vol^{11/3}} \left[ e^{-\frac{2}{\sqrt{3}}(\langle\Sigma\rangle+\sigma)}
- \frac{(2k)^{2/3}\tilde{C}\,\vol^{1/3}}{B_1|W_0|^2} e^{-\frac{1}{\sqrt{3}}(\langle\Sigma\rangle+\sigma)}+\frac{2 B_2}{B_1}\,e^{\frac{1}{\sqrt{3}}(\langle\Sigma\rangle+\sigma)}\right].
\label{Vinf2}
\ee
We now fix $\langle\Sigma\rangle$ from the requirement that this potential is minimised at the origin, $\sigma=0$, giving
\be
e^{-\frac{2\langle\Sigma\rangle}{\sqrt{3}}}-\frac{(2k)^{2/3}\tilde C\,\vol^{1/3}}{2 B_1 |W_0|^2}\, e^{-\frac{\langle\Sigma\rangle}{\sqrt{3}}}-\frac{B_2}{B_1}\,e^{\frac{\langle\Sigma\rangle}{\sqrt{3}}}=0\,,
\label{constraint}
\ee
and suggesting $e^{-\langle\Sigma\rangle/\sqrt{3}} \sim \vol^{1/3}$.  However, it is convenient to use this equation to eliminate the flux dependent constant $\tilde C$, rather than $\langle\Sigma\rangle$, from the inflaton potential, allowing us to write the latter as \cite{Cicoli:2008gp, Cicoli:2016chb}
\be 
\label{Vinf3}
V_\text{inf}(\sigma)=V_0 \left[ e^{-\frac{2\sigma}{\sqrt{3}}}-2\, e^{-\frac{\sigma}{\sqrt{3}}} +2\mathcal{R}\, \cosh \left(\frac{\sigma}{\sqrt{3}}\right) \right],
\ee
where 
\be
V_0\equiv \frac{B_1|W_0|^4}{(2k)^{1/3}\,\vol^{11/3}}\, e^{-\frac{2\langle\Sigma\rangle}{\sqrt{3}}} \qquad\text{and}\qquad \mathcal{R}\equiv \frac{2B_2}{B_1}\,e^{\sqrt{3} \langle\Sigma\rangle} = \frac{2\Pi_2}{\Pi_1}\,\frac{\langle\tau_1\rangle}{\langle\tau_2\rangle}\,.
\label{R}
\ee
The value of the inflaton potential at the minimum, $V_\text{inf}(0)= - V_0\left(1-2\mathcal{R}\right)$ is non-vanishing. However, if we re-tune the uplift $\kappa$ accordingly
\be
\kappa \to \kappa+ V_0\left(1-2 \mathcal{R}\right)\vol^n\,,
\ee 
the minimum of the inflaton potential is effectively shifted to zero, so that we now have
\be
V_\text{inf}(\sigma)=V_0 \left[  \left(1-e^{-\frac{\sigma}{\sqrt{3}}} \right)^2 -2\mathcal{R}\left(1- \cosh \left(\frac{\sigma}{\sqrt{3}}\right)\right)\right]. 
\label{Vinf4}
\ee
The slow-roll parameters can now be computed directly, giving
\begin{eqnarray}
\epsilon &=& \frac12 \left(\frac{V_\text{inf}'}{V_\text{inf}} \right)^2 =\frac23 \left[ \frac{e^{-\frac{\sigma}{\sqrt{3}}} -e^{-\frac{2\sigma}{\sqrt{3}}}+\mathcal{R}\,\sinh \left(\frac{\sigma}{\sqrt{3}}\right)}{\left(1-e^{-\frac{\sigma}{\sqrt{3}}} \right)^2-2\mathcal{R}\left(1- \cosh \left(\frac{\sigma}{\sqrt{3}}\right)\right) } \right]^2 \\
\eta &=& \frac{V_\text{inf}''}{V_\text{inf}}= - \frac23 \left[ \frac{e^{-\frac{\sigma}{\sqrt{3}}}- 2\,e^{-\frac{2\sigma}{\sqrt{3}}} -\mathcal{R}\,\cosh \left(\frac{\sigma}{\sqrt{3}}\right)}{\left(1-e^{-\frac{\sigma}{\sqrt{3}}} \right)^2-2\mathcal{R}\left(1- \cosh \left(\frac{\sigma}{\sqrt{3}}\right)\right) } \right].
\end{eqnarray}
When $\epsilon \ll 1$ and $|\eta| \ll 1$, these feed into the standard observables corresponding to the amplitude of scalar perturbations, $\mathcal{A}_s$, the corresponding spectral index, $n_s$, and the tensor to scalar ratio, $r$:
\be
\mathcal{A}_s=\frac{V_\text{inf}}{24 \pi^2 \epsilon}\,, \qquad n_s=1+2\eta-6 \epsilon\,, \qquad r=16\,\epsilon\,.
\ee

\begin{figure}[ht]
\begin{center}
\includegraphics[width=0.6\textwidth]{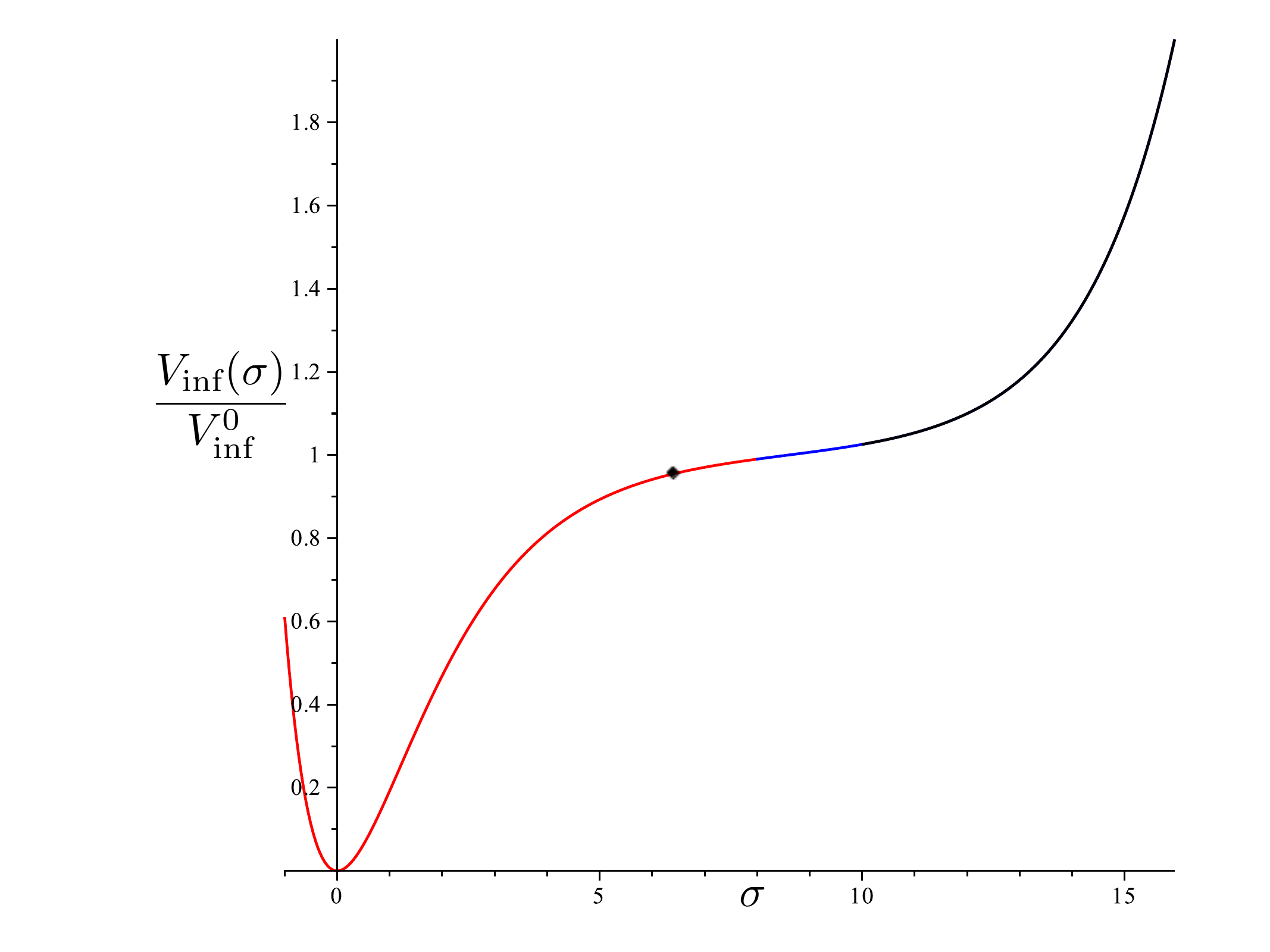}
\caption{Form of the inflaton potential \eqref{Vinf3} for $\mathcal{R}=0.0001$. The black region corresponds to the asymptotic regime with $e^{-\frac{\sigma}{\sqrt{3}}} \ll \mathcal{R} \ll 1$. The blue region is the intermediate regime with $e^{-\frac{2\sigma}{\sqrt{3}}} \ll \mathcal{R} \ll e^{-\frac{\sigma}{\sqrt{3}}} \ll 1$. The red region is the regime with $\mathcal{R} \ll e^{-\frac{2\sigma}{\sqrt{3}}} \ll 1$. The black diamond corresponds to the point where $e^{-\frac{\sigma}{\sqrt{3}}} \simeq 0.0246$, yielding observationally viable inflation.}
\label{fig:inflation}
\end{center}
\end{figure}

The best fit of Fibre Inflation to cosmological data has been studied in \cite{Cicoli:2020bao}. Here we just note that for $0\leq \mathcal{R} \ll 1$, which can be achieved for $0\leq \Pi_2\ll \Pi_1$ and/or $\langle\tau_1\rangle\ll\langle\tau_2\rangle$, we can consider three regimes for large $\sigma$, as shown in Fig. \ref{fig:inflation}. The first is the asymptotic regime for which $e^{-\frac{\sigma}{\sqrt{3}}} \ll \mathcal{R} \ll 1$ where $\epsilon \simeq 1/6$ and $\eta \simeq 1/3$. The corresponding spectral index, $n_s\simeq 2/3$, violates the observational constraint, $n_s=0.9649\pm 0.0042$ \cite{Planck:2018jri}. Moving the inflaton inwards, we now consider the intermediate regime, for which $e^{-\frac{2\sigma}{\sqrt{3}}} \ll \mathcal{R} \ll e^{-\frac{\sigma}{\sqrt{3}}} \ll 1$. This gives $\epsilon\simeq\frac32 \eta^2$ and $\eta\simeq\frac13 \mathcal{R}\, e^{\frac{\sigma}{\sqrt{3}}}>0$, resulting in a blue spectrum for scalar perturbations, again ruled out by observations \cite{Planck:2018jri}. The final regime of interest corresponds to an even smaller value of the inflaton, with $0\leq \mathcal{R} \ll e^{-\frac{2\sigma}{\sqrt{3}}} \ll 1$. This also gives $\epsilon\simeq\frac32 \eta^2$ although now with $\eta\simeq-\frac23 \,e^{-\frac{\sigma}{\sqrt{3}}}$. This yields a red spectrum for scalar perturbations that can be compatible with data. We will say more about the phenomenological constraints for this viable regime in Sec. \ref{sec:constraints}. For now, we note that once inflation ends, the inflaton will oscillate about the minimum of its potential with square mass
\be
m^2_\sigma=\frac23\, V_0 \left(1+\mathcal{R}\right) \simeq \frac23\,V_0\,.
\label{msigma}
\ee
The perturbative decay of the inflaton then leads to a radiation dominated era. The details of the reheating process depend on the underlying brane construction which realises the Standard Model. There are three possibilities: ($i$) the SM lives on D7-branes wrapped around the inflaton \cite{Cicoli:2018cgu}; ($ii$) the SM is on D7-branes wrapped around a blow-up mode in the geometric regime \cite{Cicoli:2023njy}; ($iii$) the SM is on D3-branes at a Calabi-Yau singularity \cite{Cicoli:2022uqa}. In all cases, the resulting number of efoldings of inflation is around $N_e\simeq 52$, which would correspond to $e^{-\frac{\sigma}{\sqrt{3}}} \simeq 0.0246$ for $\mathcal{R}\simeq 0.0001$. On top of SM particles, the inflaton decay produces also relativistic bulk closed string axions which behave as dark radiation. Their contribution to effective number of neutrino species, $N_{\rm eff}$, is however tamed by the inflaton decay into SM gauge bosons and Higgses. Interestingly, due to the high supersymmetry breaking scale (even in the presence of sequestering for D3-branes), if stable, neutralinos would always overproduce dark matter. Hence $R$-parity has necessarily to be broken, allowing these heavy modes to decay \cite{Allahverdi:2023nov}. Dark matter could instead be given by primordial black holes \cite{Cicoli:2018asa, Cicoli:2022sih}, or by the QCD axion realised as the phase of an open string mode on D3-branes \cite{Cicoli:2022uqa}. In the latter case, the QCD axion could have a decay constant below the inflationary Hubble scale, allowing it to avoid isocurvature bounds. On the other hand, as we shall see in Sec. \ref{DMsec} and App. \ref{DMsecnew}, fuzzy dark matter produced via the misalignment mechanism is not a viable option here since the contribution of ultra-light bulk axions to dark matter is negligible given that they either start oscillating after matter-radiation equality, or their abundance is bounded by a combination of isocurvature and theory constraints.

\subsection{The late universe}
\label{sec:late}

We now fast forward almost 14 billion years to the modern day and consider the dynamics of the late universe. Thanks to the hierarchical change in scale, the late time behaviour stems from non-perturbative corrections. With all saxions now stabilised along with the axion $\theta_s$, the late time scalar potential \eqref{Vlate} is given by
\be
V_\text{late}= \Lambda_2^4\left[1-\cos\left(a_2 \theta_2\right)\right] + \Lambda_1^4\left[1-\cos\left(a_2\theta_2+a_1\theta_1\right)\right]
\label{Vlate2}
\ee
where we have adjusted the uplifting contribution to obtain a Minkowski vacuum and we have defined
\begin{equation}
\Lambda_2^4 \equiv  \frac{4 |W_0| A_2}{\vol^2} \,a_2\langle\tau_2\rangle\,e^{-a_2\langle\tau_2\rangle} \gg \Lambda_1^4 \equiv \Lambda_2^4 \left(1+\frac{a_1 \langle\tau_1\rangle}{a_2\langle\tau_2\rangle}\right)A_1\,e^{-a_1\langle\tau_1\rangle}\,.
\label{LamI}
\end{equation}
The contribution proportional to $\Lambda_1^4$ is subleading with respect to the one proportional to $\Lambda_2^4$ due to the additional $e^{-a_1\langle\tau_1\rangle}$ suppression factor (for natural $\mathcal{O}(1)$ values of $A_1$). 

The kinetic terms for the two dynamical axions at late times are given by
\be
\mathcal{L}_\text{kin} \supset -\frac{1}{4\langle\tau_1\rangle^2} (\del \theta_1)^2 -\frac{1}{2\langle\tau_2\rangle^2} (\del \theta_2)^2\,.
\ee
Since $\tau_1$ and $\tau_2$ are stabilised at late times, we can readily introduce the canonical fields:
\be
\theta_1=\sqrt{2}\,\langle\tau_1\rangle\, \phi_1\qquad\text{and}\qquad\theta_2=\langle\tau_2\rangle\, \phi_2\,.
\ee
In terms of the canonical fields, the late time potential \eqref{Vlate2} takes the form 
\be
V_\text{late}\simeq  \Lambda_2^4\left[1-\cos \left(\frac{\phi_2}{f_2} \right)\right]+\Lambda_1^4\left[1-\cos \left(\frac{\phi_1}{f_1}  + \frac{\phi_2}{f_2} \right)\right]
\label{Vlate3}
\ee
where
\begin{equation}
f_1 \equiv \frac{N_1}{2\sqrt{2}\pi\langle\tau_1\rangle}\qquad\text{and}\qquad f_2 \equiv \frac{N_2}{2\pi\langle\tau_2\rangle}\,.
\label{DecayConstants}
\end{equation}
In Sec. \ref{sec:constraints} these expressions for the decay constants will allow us to easily compare with observational constraints on ultra-light axions.

As we will see in a moment, at leading order, this potential depends just on $\phi_2$ and the axion $\phi_1$ is an exactly flat direction. Hence $\phi_1$ behaves as a natural candidate to drive dynamical dark energy once the subleading contribution is included. Quintessence can be realised around the saddle point where the $\phi_2$ direction is stable while $\phi_1$ is around the maximum of its effective potential. 

Recalling that $\Lambda_2 \gg \Lambda_1$, the mass hierarchy around this saddle point is
\be
m_2\simeq \frac{\Lambda_2^2}{f_2}\gg |m_1| \simeq \frac{\Lambda_1^2}{f_1}\,.
\label{maxions}
\ee
Due to this hierarchy, the leading order potential looks like
\begin{equation}
V(\phi_2)\simeq \Lambda_2^4\left[1-\cos\left(\frac{\phi_2}{f_2}\right)\right]
\label{Veffh}
\end{equation}
The heavy mode $\phi_2$ can therefore be integrated out by setting it to its minimum at $\langle\phi_2\rangle = 0$. The effective potential for the light axion $\phi_1$ then becomes
\begin{equation}
V(\phi_1)\simeq \Lambda_1^4\left[1-\cos\left(\frac{\phi_1}{f_1}\right)\right]
\label{Veff}
\end{equation}
In this scenario, the heavier axion $\phi_2$ is stable and oscillates about equilibrium point, in principle providing a possible candidate for dark matter today, although as we will see in Sec. \ref{sec:constraints}, its abundance is tiny. On the other hand, the light axion $\phi_1$ is displaced from its minimum and plays the role of the quintessence field. The corresponding slow-roll parameters go as $\epsilon \simeq |\eta| \simeq 1/f_1^2$. Note that $\eta$ is small in absolute value, and so the light axion can be identified with a quintessence field in slow-roll.

\section{Phenomenological constraints}
\label{sec:constraints}

Having described our setup and the corresponding dynamics for volume stabilisation, inflation and late universe cosmology, we now impose a number of phenomenological constraints on our model in order to pin down a consistent set of parameters and to extract any important predictions.

\subsection{Mass spectrum}

The string scale is given by
\be
M_s = \frac{g_s^{1/4}\,M_p}{\sqrt{4 \pi \vol}}\,.
\ee
This is lower than the Planck scale in the limit where the effective field theory is under control, i.e. for perturbative strings $g_s\ll 1$ at large volume $\vol\gg1$. Further, the Kaluza-Klein scale  associated with the (isotropic) compact extra dimensions is given by 
\be
M_\text{KK}\simeq \frac{M_p}{\sqrt{4 \pi} \,\vol^{2/3}}\,.
\ee
Both the Kaluza-Klein and the string scale should exceed the scale of all the moduli masses in order to trust the low-energy effective theory described in Sec. \ref{sec:setup}. An important reference scale for the moduli mass spectrum is the gravitino mass
\be
m_{3/2}\simeq \frac{|W_0|\,M_p}{\mathcal{V}}\,.
\ee
In order to be consistently decoupled during the early accelerated expansion, all moduli associated with the stabilisation of the volume should have masses above the scale of inflation:
\be
H_\text{inf} \simeq\sqrt{\frac{V_\text{inf}}{3 M_p^2}}\simeq\sqrt{\frac{V_0}{3 M_p^2}}\,.
\ee
Since the overall volume is much larger than one in string units and the string coupling is small, it follows that
\be
m_{\theta_s} \sim m_{\tau_s} \sim m_{3/2}\gg m_\vol > m_\sigma\sim H_\text{inf}\gg m_2\gg m_1\sim H_0\,,
\ee
where the masses of the moduli and the axions are given by \eqref{axionmass}, \eqref{saxionmasses}, (\ref{msigma}) and (\ref{maxions}). To safely comply with constraints from the muon anomaly, the gravitino mass is bounded from below by $m_{3/2} \gtrsim 0.1$ meV \cite{Ferrer:1997yz}. This is automatically satisfied in the present model since in Sec. \ref{InflConstr} we will see that inflationary constraints fix $\mathcal{V}\simeq 10^3$ which, for $|W_0|\sim\mathcal{O}(1)$, gives $m_{3/2}\simeq 10^{15}$ GeV. The scale of supersymmetry breaking is therefore very large. Even in a sequestered scenario with the MSSM realised on D3-branes \cite{Aparicio:2014wxa}, the scale of the soft terms would be very high, $M_{\rm soft}\simeq 10^{10}$-$10^{11}$ GeV. 

\subsection{Inflationary constraints} 
\label{InflConstr}

We begin with the  inflationary observables. The amplitude for scalar perturbations, the corresponding spectral index and the tensor-to-scalar ratio are given by \cite{Planck:2018jri}
\be
\label{infcon}
\mathcal{A}_s=\frac{V_\text{inf}}{24\pi^2  M_p^4  \epsilon}  \simeq 2.09 \times 10^{-9}\,, \qquad n_s=1+2\eta-6 \epsilon\simeq  0.9649\,, \qquad r=16 \epsilon<0.056\,.
\ee
At the end of Sec. \ref{sec:inflation} we presented a viable inflationary scenario with $\mathcal{R} \ll e^{-\frac{2\sigma}{\sqrt{3}}} \ll 1$ throughout the observably relevant range, where $\sigma$ is the canonical inflaton field and $\mathcal{R}$ is given by \eqref{R}. This gave $\epsilon \simeq \frac32 \eta^2 \ll |\eta|$, with $\eta \simeq -\frac23\, e^{-\frac{\sigma}{\sqrt{3}}}$, with the following predictions for the two main cosmological observables: 
\be
n_s\simeq 0.9672\qquad\text{and}\qquad r\simeq 0.0079
\ee
for $\mathcal{R}=0.0001$ and $N_e\simeq 52$ (corresponding to $e^{-\frac{\sigma}{\sqrt{3}}}  \simeq 0.0246$), as given by the study of the post-inflationary evolution of our model. The scalar spectral index is in perfect agreement with data and the predicted value of the tensor-to-scalar ratio is at the edge of detection. 

Recall that the inflaton is given in terms of the ratio of the large K\"ahler saxions, as $\sigma=\ln (\tau_1/\tau_2)/\sqrt{3}-\langle\Sigma\rangle$. In principle, the factor of $e^{\frac{\langle\Sigma\rangle}{\sqrt{3}}}$ can be obtained as a root of \eqref{constraint}. To find that root, we rewrite that equation as
\be
1-\frac{\mathcal{R}}{2}-\frac{(2k)^{2/3} \tilde{C}\, \vol^{1/3}}{2 B_1 |W_0|^2}\,e^{\frac{\langle\Sigma\rangle}{\sqrt{3}}}=0\,.
\ee
For $\mathcal{R}\ll 1$ and using $\vol \simeq (2k)^{-1/2} \sqrt{\tau_1} \tau_2$, this equation is now easily solved to a good approximation as
\be
e^{\frac{\langle\Sigma\rangle}{\sqrt{3}}} \simeq \frac{2 B_1 |W_0|^2}{(2k)^{2/3}\tilde{C}\, \vol^{1/3}}\qquad\Leftrightarrow\qquad \langle\tau_1\rangle \simeq \frac{2}{k}\left(\frac{B_1 |W_0|^2}{\tilde{C}}\right)^2\,,
\label{tau1vev}
\ee
where $\langle\tau_1\rangle$ denotes the value of $\tau_1$ at the post-inflationary minimum. During inflation, $\mathcal{V}$ is fixed while $\tau_1$ changes. At CMB horizon exit at $N_e\simeq 52$, one obtains
\be
\tau_1 = \langle\tau_1\rangle\,e^{\frac{2}{\sqrt{3}}\sigma}\simeq 1650 \, \langle\tau_1\rangle\,.
\label{Tau1he}
\ee
We have now collected all the relevant information to derive two constraints on the UV parameters of our model from inflation:
\begin{enumerate}
\item \textbf{Amplitude of density perturbations:} For the field value (\ref{Tau1he}), $\epsilon \simeq 0.0005$ which, when combined with the scale of scalar perturbations \eqref{infcon}, gives the scale of inflation to be
\be
H_\text{inf} =\sqrt{\frac{V_\text{inf}}{3 M_p^2}}\simeq 0.9 \times  10^{-5}\, M_p \simeq 2\times 10^{13}\,\text{GeV}\,.
\label{Hinf}
\ee
In order to relate this to the parameters of our model, we evaluate $V_{\rm inf}$ at CMB horizon exit with $V_0$ given in \eqref{R} combined with (\ref{tau1vev}), obtaining a constraint on the volume
\be
\vol \simeq 1250 \left(\frac{k\,\tilde C^2}{B_1} \right)^{1/3}
\label{Cond1}
\ee

\item \textbf{Taming of steepening corrections}: As we have already stressed, the inflationary potential (\ref{Vinf4}) can yield enough efoldings of slow-roll inflation if the positive exponential term inside the $\cosh$ is multiplied by a small prefactor. If the topology of the $\tau_2$ divisor is such that $\Pi_2=0$ \cite{Cicoli:2023njy}, the prefactor $\mathcal{R}$ actually vanishes since (\ref{R}) gives $\mathcal{R}=0$. However, here and in Sec \ref{sec:inflation} we have focused on a numerical example with $\mathcal{R}\simeq 0.0001$ which clearly requires $\Pi_2>0$. Recalling that $\Pi_1=24$ for a K3 divisor, we obtain the following relation between the two large saxions:
\be
\langle\tau_1\rangle\simeq \frac{0.0012}{\Pi_2}\,\langle\tau_2\rangle
\label{Cond2}
\ee
\end{enumerate}

\subsection{Dark energy constraints}

We now consider how the parameters of our model affect the details of its late time cosmological behaviour. Reproducing the correct dark energy dynamics imposes two further constraints:
\begin{enumerate}
\item \textbf{Accelerating solution:} As discussed in \cite{Cicoli:2021skd}, axion hilltop quintessence with sub-Planckian decay constant is only viable if the axion stays close to the  corresponding hilltop. Due to quantum diffusion during inflation, the light fields will typically be displaced by $\Delta \phi\sim H_{\rm inf}$ from their initial position. We introduce $\Delta_{\max}(f_1)$, which  measures the maximum allowed displacement of the axion from the hilltop compatible with current bounds on the dark energy equation of state parameter for a given decay constant $f_1$. Generically for $\Delta_{\max}(f_1)>H_\text{inf} \simeq 10^{-5} M_p$ the initial conditions for the dark energy axion are not spoilt by quantum diffusion during inflation. As can be seen from Fig. \ref{fig:f_vs_Phi0} (where $f$ should be identified with $f_1$), this implies that for the present model $f_1\gtrsim 0.08\, M_p$. Let us therefore set:
\be
f_1\simeq 0.085\, M_p\quad\Leftrightarrow\quad \langle\tau_1\rangle\simeq 1.324\,N_1
\label{Cond3}
\ee

\begin{figure}[ht]
\begin{center}
\includegraphics[width=0.7\textwidth]{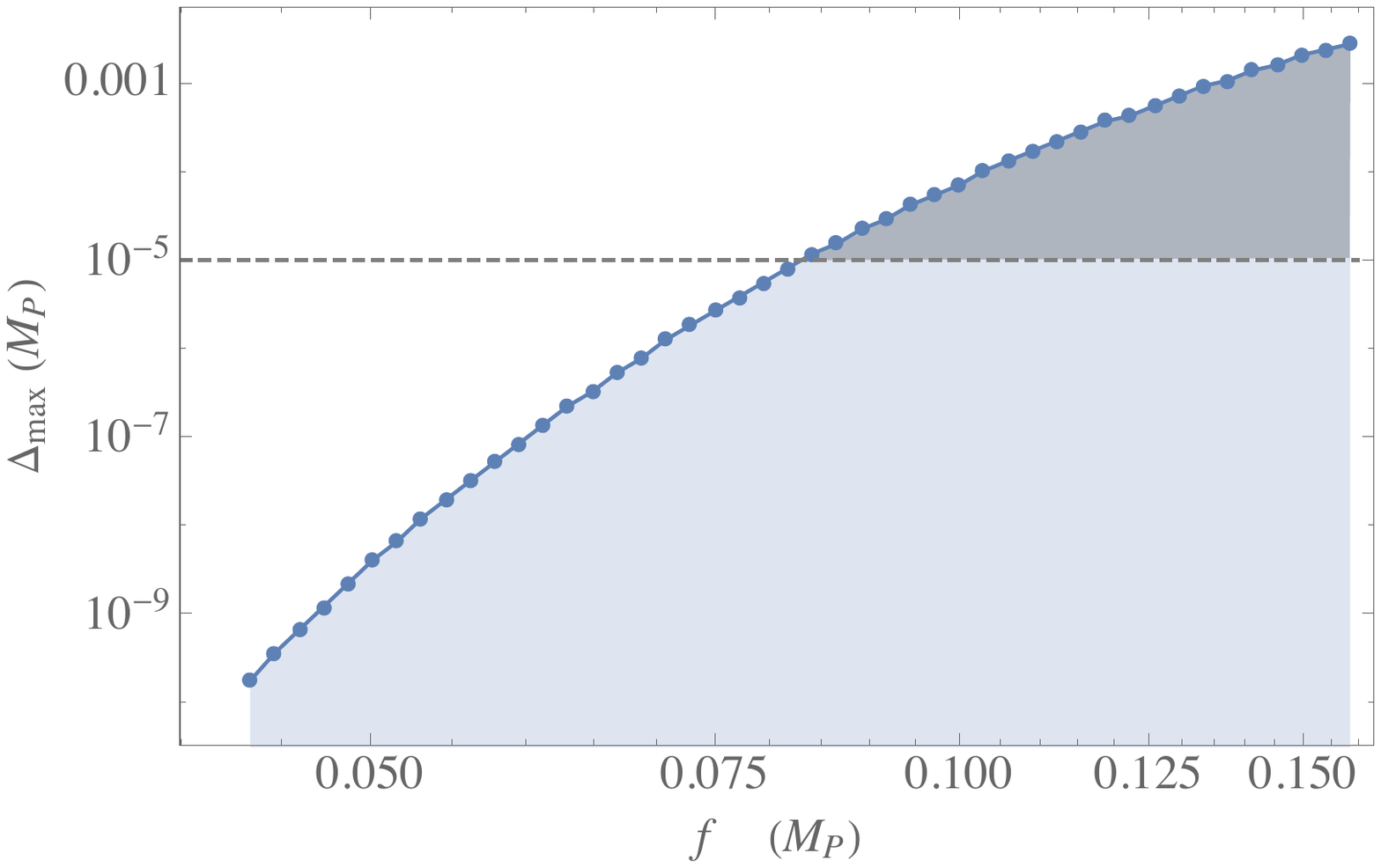}
\caption{Maximum displacement from the hilltop compatible with bounds on the dark energy equation of state parameter as a function of the decay constant $f$. The dashed line corresponds to the scale of inflation in the present model, $H_{\rm inf}\simeq 10^{-5}\,M_p$, while the grey region denotes the viable parameter range: $\Delta_{\max}(f)>H_\text{inf}$.}
\label{fig:f_vs_Phi0}
\end{center}
\end{figure}

\item \textbf{Dark energy scale:} The scale of dark energy is set by the size of the late time potential at the corresponding saddle, giving
\be
\Lambda_1^4 \simeq 10^{-120}\,M_p^4
\ee
Setting in (\ref{LamI}) natural $\mathcal{O}(1)$ values of the UV parameters as $|W_0|=A_1=A_2=1$, together with $f_1\simeq 0.085\,M_p$ from (\ref{Cond3}) and $\vol\simeq 900$ in compatibility with (\ref{Cond1}), this gives:
\be
f_2\simeq 0.0038\,M_p\quad\Leftrightarrow\quad \langle\tau_2\rangle\simeq 41.56\,N_2\,.
\label{Cond4}
\ee

Therefore the masses of the two ultra-light axions from (\ref{maxions}) become:
\begin{eqnarray}
m_2&\simeq& \frac{\Lambda_2^2}{f_2}\simeq 1.8\times 10^{-56}\,M_p\simeq 4.5\times 10^{-29}\,{\rm eV} \nonumber \\
|m_1| &\simeq& \frac{\Lambda_1^2}{f_1}\lesssim 1.2\times 10^{-59}\,M_p \simeq 2.9\times 10^{-32}\,{\rm eV}\,.
\end{eqnarray}
\end{enumerate}

\subsection{Dark matter constraints}
\label{DMsec}

Given that $m_2\simeq 10^{-29}$ eV, the heavier axion $\phi_2$ starts oscillating after matter-radiation equality which occurs at a Hubble scale of order $H_\text{eq} \simeq 10^{-54}\,M_p\simeq 10^{-27}$ eV. Hence $\phi_2$ cannot compose all the dark matter. In fact, its contribution to dark matter via the misalignment mechanism is \cite{Marsh:2015xka}:
\be 
\frac{\Omega_2}{\Omega_m}\simeq \frac32 \left(\frac{\delta \phi_2}{M_p}\right)^2
\label{abundance}
\ee
Note that we expect
\be
\frac{H_\text{inf}}{2\pi}\simeq 1.4\times 10^{-6} \lesssim \frac{\delta \phi_2}{M_p} \lesssim  \pi f_2 \lesssim 0.012
\ee
with the lower bound coming from quantum diffusion during inflation \cite{Hardwick:2017fjo,Cicoli:2021skd}. Plugging this in (\ref{abundance}), we obtain the following upper bound for the axion dark matter fraction:
\be 
\frac{\Omega_2}{\Omega_m}\lesssim 0.0002
\ee
which implies a negligible contribution to the dark matter abundance. 

Given that $f_2 > H_{\rm inf}$, the Peccei-Quinn symmetry for $\phi_2$ is broken during inflation, and so the abundance of axion dark matter is subject to isocurvature constraints. In fact, the axion $\phi_2$ generates isocurvature perturbations whose amplitude goes as \cite{Marsh:2015xka}:
\be
\mathcal{A}_I =\left(\frac{\Omega_2}{\Omega_m}\right)^2 \left(\frac{H_\text{inf}}{\pi \delta \phi_2}\right)^2\simeq \frac94\left(\frac{H_\text{inf}}{\pi M_p}\right)^2\left(\frac{\delta \phi_2}{M_p}\right)^2\lesssim 2.7\times 10^{-15}\quad\Rightarrow\quad\frac{\mathcal{A}_I}{\mathcal{A}_s}\lesssim 1.3\times 10^{-6}\,,
\label{AI}
\ee
which is perfectly in agreement with the current observational bound 
$\mathcal{A}_I/\mathcal{A}_s<0.038$ \cite{Marsh:2015xka}, with $A_s \simeq 2.09 \times 10^{-9}$.

\subsection{A numerical example}

Combining the two inflationary constraints (\ref{Cond1}) and (\ref{Cond2}) with the two dark energy constraints (\ref{Cond3}) and (\ref{Cond4}), we can fix the four quantities $\langle\tau_1\rangle$, $\langle\tau_2\rangle$, $N_1$ and $N_2$ in terms of the underlying parameters $k$, $\Pi_2$, $\tilde{C}$ and $B_1$. 

Setting $\Pi_2=1$, one can write (\ref{Cond2}) in terms of (\ref{Cond3}) and (\ref{Cond4}), finding:
\be
\frac{\langle\tau_1\rangle}{\langle\tau_2\rangle}\simeq 0.032\,\frac{N_1}{N_2}\simeq 0.0012\quad\Leftrightarrow\quad \frac{N_1}{N_2}\simeq 0.038
\label{Rcond}
\ee
The lowest natural numbers which satisfy this condition are $N_1=1$ (as for a typical poly-instanton correction) and $N_2=27$. Plugging these numbers in (\ref{Cond3}) and (\ref{Cond4}), we obtain:
\be
\langle\tau_1\rangle\simeq 1.324\quad\text{and}\quad
\langle\tau_2\rangle\simeq 41.56\,N_2\simeq 1122\,.
\label{Cond4new}
\ee
These values can be plugged in the CY volume (\ref{CYvol}) with $k=1$ (all known fibred CY examples have $k\sim\mathcal{O}(1-10)$ \cite{Cicoli:2011it,Cicoli:2016xae,Cicoli:2017axo,Bera:2024sbx}), $\hat{k}=1$ and $\langle\tau_s\rangle$ given in (\ref{sols}) with $g_s=0.1$ and $\xi=0.5$ (as for a typical CY threefold), obtaining $\vol \simeq 900$. This value of the internal volume has to be reproduced by (\ref{Cond1}), implying:
\be
B_1 \simeq 2.65\, \tilde{C}^2
\label{Cond1new}
\ee
Moreover $\langle\tau_1\rangle\simeq 1.324$ should also match (\ref{tau1vev}) combined with (\ref{Cond1new}), giving:
\be
\tilde{C}^2 \simeq 0.095\,|W_0|^{-4}\,,
\label{tau1vevnew}
\ee
For $|W_0|=1$, these two conditions imply $\tilde{C}\simeq 0.3$ and $B_1\simeq 0.25$ which can be obtained for $|\lambda|\simeq 3.3\times 10^{-4}$ that is of the expected order of magnitude \cite{Cicoli:2023njy}.

One may wonder whether $\langle\tau_1\rangle\simeq 1.324$ is large enough to trust the effective field theory. We shall now argue that this is indeed the case. In fact, the two main conditions which we have to check for a relatively small $\langle\tau_1\rangle$ are the following:
\begin{enumerate}
\item Stringy effects can be neglected when all 2-cycle volumes are much larger than the string scale, corresponding to \cite{AbdusSalam:2020ywo}:
\be
|t_i|\gg \frac{1}{(2\pi)^2\sqrt{g_s}}\simeq 0.08\qquad \text{for}\qquad g_s=0.1\qquad\forall i=1,2,3\,.
\ee
In our case the most relevant condition is the one for $i=2$ which becomes (for $k=1)$:
\be
\langle t_2\rangle = \sqrt{2\langle\tau_1\rangle} \simeq 1.63 \gg 0.08\,.
\ee
and so stringy effects can be safely neglected.

\item String loop corrections to $K$ are subdominant with respect to $\alpha'^3$ effects when:
\be
\frac{|V_{\rm loop}|}{V_{\alpha'^3}}\ll 1\,.
\ee
In our case this condition is also easily met since for our parameter choice:
\be
\frac{|V_{\rm loop}|}{V_{\alpha'^3}}= \frac{4\sqrt{2k}\,\tilde{C}g_s^{3/2}}{3\xi\sqrt{\langle\tau_1\rangle}}\simeq 0.032\ll 1\,.
\ee
\end{enumerate}
This parameter choice is characterised by $\mathcal{O}(1)$ values of all microscopic parameters except for $N_2$ which has to be at least $27$.  This is required by the $\mathcal{R}\simeq 0.0001$ condition (\ref{Rcond}). However this condition can be relaxed if the $\tau_2$ divisor is such that $\Pi_2=0$ since one would automatically obtain $\mathcal{R}=0$. As studied in \cite{Cicoli:2023njy}, there are a few divisors with this property: dP$_3$, $\mathbb{T}^4$ and divisors with Wilson lines. In this case, $N_2$ can be smaller. For example $N_1=N_2=10$ would lead to $\vol\simeq 1056$, $\langle\tau_2\rangle\simeq 415.6$ and $\langle\tau_1\rangle\simeq 13.42$ which improves the control over the effective field theory.

\section{Conclusions} 
\label{sec:conc}

Explaining cosmic acceleration compatible with observations in the early universe and today is still a formidable challenge. As far as acceleration in the late universe is concerned, the simplest explanation seems to rely on a de Sitter vacuum, even if very hard to achieve explicitly. In fact, dynamical dark energy models possess additional problems, such as unobserved fifth forces, perturbative stability of the quintessence potential, and obtaining the right cosmological constant scale without too much lowering of the mass spectrum for string and Kaluza-Klein modes, supersymmetric particles and moduli below the TeV scale. However, future cosmological observations might point towards a dynamical dark energy model, not least in light of the tantalising results of the DESI collaboration \cite{DESI:2024mwx}. It is therefore crucial to build quintessence models that  can match data and be embedded in a consistent UV theory like string theory.

In this paper we made progress in this direction by presenting a string model which can describe the history of the universe from inflation to quintessence within the framework of a 4D effective field theory where all the moduli are stabilised and the corrections to the leading order results are under control. More precisely, we focused on type IIB Calabi-Yau flux compactifications where moduli stabilisation is best understood. Cosmic acceleration in the early universe is realised as in Fibre Inflation where the inflaton is a fibration K\"ahler modulus with a potential generated by perturbative corrections to the K\"ahler potential. The inflationary scale is around $H_{\rm inf}\simeq 10^{13}$ GeV. 

The model also features two ultra-light axions that  behave as spectator fields during inflation and acquire isocurvature perturbations. Reheating is driven by the perturbative decay of the inflaton field. This decay, on top of damping energy into the Standard Model, also produces relativistic axions that contribute to dark radiation. The exact amount of dark radiation production depends on the details of the D-brane realisation of the Standard Model. Dark matter can arise from primordial black holes, but also from an open string QCD axion when the Standard Model lives on D3-branes. 

The scale of the potential of the two ultra-light axions is naturally very suppressed due to the smallness of non-perturbative effects. Therefore, the current Hubble scale can in principle be easily achieved without  need for severe fine tuning. However, the axionic potential is generically too steep to sustain an accelerating solution, and so quintessence can only be realised  close to the maximum. Quantum diffusion during inflation would then tend to ruin this realisation of quintessence by pushing the axion away from this region. For Fibre Inflation with $H_{\rm inf}\simeq 10^{13}$ GeV, this does not happen only if one of the two ultra-light axions has a decay constant of order $f\simeq 0.1\,M_p$. However, if this axion receives a mass through standard non-perturbative effects, the corresponding instanton action would be too small to reproduce the current dark energy scale. We therefore proposed a scenario based on tiny poly-instantons corrections to the superpotential which can indeed be generated by a K3 divisor \cite{Lust:2013kt} and, above all, can lead to a very large suppression, allowing us to match the cosmological constant scale even for $f\simeq 0.1\,M_p$. As a result,  the lighter axion drives quintessence, while the heavier one receives mass by gaugino condensation. This heavy axion behaves as a spectator field which starts oscillating after matter-radiation equality, and as such,  gives only a negligible contribution to dark matter.

As outlined in detail in App. \ref{AppAlignment}, we also considered vacua where the decay constants of both axions are around $0.01\,M_p$. In this case, quantum diffusion during inflation would ruin the initial conditions to realise quintessence. We therefore implemented a mild alignment mechanism which yields a mass hierarchy between the two ultra-light axions. The lighter axion has a mass of order $10^{-32}$ eV and an effective decay constant of order $0.1\,M_p$ which is large enough to yield a quintessence model compatible with quantum diffusion during inflation. The heavier axion is stabilised during quintessence and behaves as a spectator field. It cannot yield a large contribution to dark matter via the misalignment mechanism due to isocurvature perturbation bounds and theory constraints from the underlying UV embedding. In fact, this field can constitute at most  $5\%$ of dark matter when its mass is around $10^{-23}$ eV. Interestingly, our model can satisfy the weak gravity conjecture applied to axion physics thanks to the inclusion of an extra instanton contribution to the superpotential.

All our models are realised within a string-inspired 4D supergravity theory. It has all the ingredients of a proper string compactification, except for an explicit description of the underlying Calabi-Yau orientifold, brane setup compatible with D3 and D7 tadpole cancellation, and flux stabilisation of the dilaton and complex structure moduli. A step forward in this direction would be crucial to realise a complete string model that is able to describe the history of the universe from inflation to quintessence. We leave this task for future work.

\section*{Acknowledgments}

This work is partly supported by the COST (European Cooperation in Science and Technology) Action COSMIC WISPers CA21106.

\appendix

\section{Quintessence from alignment}
\label{AppAlignment}

If the underlying parameters which determine $\langle\tau_1\rangle$ from (\ref{tau1vev}) are such that the condition (\ref{Cond3}) is not satisfied by any $N_1$, one might naively think that the quintessence solution is necessarily destroyed by quantum diffusion during inflation. However this is not the case in the presence of axion alignment which we analyse in this appendix.

\subsection{An alternative setup}
\label{sec:setupew}

The mechanism of axion alignment requires a slight modification of the non-perturbative superpotential that now becomes:
\be
W_{\rm np}=A_s\,e^{-a_s T_s}+\sum_{i=1}^3 A_i \,e^{-a_i\left(q_{i1} \,T_1 + q_{i2}\,T_2\right)}
\ee
The winding numbers $q_{ij}$, with $i=1,2,3$ and $j=1,2$, are integers labelling the number of times the $i$-th brane stack winds around the cycle $T_j$ \cite{2106.09853}. Here we consider up to four stacks: one required to implement moduli stabilisation in LVS, wrapping only the small cycle modulus $T_s$; two more wrapping only the large cycle moduli, $T_1$ and $T_2$, giving rise to a rich axion phenomenology at late times; and finally, a third stack wrapping $T_1$ and $T_2$. This stack will be suppressed by a large instanton action, and so will not affect the cosmological dynamics. However, as we will see, its inclusion is needed to be compatible with the axionic weak gravity conjecture \cite{Rudelius:2015xta, 2106.09853}.

The late time potential becomes
\be 
V_\text{late}= - \sum_{i=1}^3 \frac{4 |W_0| A_i}{\vol^2}\, a_i \left(q_{i1} \,\tau_1 + q_{i2}\,\tau_2 \right)\,e^{-a_i\left(q_{i1} \,\tau_1 + q_{i2}\,\tau_2 \right)} \cos (a_i q_{i1} \,\theta_1 + a_i q_{i2}\,\theta_2)
\label{Vlatenew}
\ee
where we have again assumed, without loss of generality, that $A_i$ for $i=1,2,3$ are all real and positive. We have kept the leading order terms and dropped those that are relatively suppressed in the large volume limit. Although not the focus here, we note in passing that we can recover the mathematics of the poly-instanton scenario discussed in the main text by replacing 
\be
A_1 \to A_1 A_2\,, \quad  q_{11} \to 1\,,  \quad q_{12} \to  N_1/N_2\,,  \quad q_{21} \to 0\,, \quad q_{22} \to 1\,\quad A_3\to 0\,.
\ee

\subsection{Quintessence potential}
\label{sec:latenew}

After stabilising all saxions and the axion $\theta_s$, the late time scalar potential \eqref{Vlatenew}  is given by
\be
V_\text{late}= \sum_{i=1}^3  \hat{\Lambda}_i^4\left[1-\cos (a_i q_{i1} \,\theta_1 + a_i q_{i2}\,\theta_2)\right]
\label{Vlate2new}
\ee
where we have adjusted the uplifting contribution to obtain a Minkowski vacuum and we have defined
\begin{equation}
\hat{\Lambda}_i^4 \equiv  \frac{4 |W_0| A_i}{\vol^2}\, a_i \left(q_{i1} \,\tau_1 + q_{i2}\,\tau_2 \right)\,e^{-a_i\left(q_{i1} \,\tau_1 + q_{i2}\,\tau_2 \right)} 
\label{LamInew}
\end{equation}
 A third stack of branes (corresponding to $i=3$) has only been included to ensure compatibility with the axion weak gravity conjecture, and should not have a significant impact on the late time dynamics. This suggests that $\hat{\Lambda}_3$ has to be suppressed relative to $\hat{\Lambda}_1$ and $\hat{\Lambda}_2$, something that can be achieved with a suitable choice of $a_3$, $q_{31}$ and $q_{32}$ \cite{Rudelius:2015xta, 2106.09853}. We will return to this point in Sec. \ref{sec:constraintsnew} - for now, we assume that the $i=3$ contribution to the late time potential can be neglected and restrict the sum to run from $1$ to $2$.

In terms of the canonical fields, the late time potential \eqref{Vlate2new} takes the form (ignoring the subleading $i=3$ term)
\be
V_\text{late}\simeq  \hat{\Lambda}_1^4\left[1-\cos \left(\frac{q_{11}}{N_1} \,\frac{\phi_1}{\mathfrak{f}_1} + \frac{q_{12}}{N_1}\,\frac{\phi_2}{\mathfrak{f}_2} \right)\right]+\hat{\Lambda}_2^4\left[1-\cos \left(\frac{q_{21}}{N_2} \,\frac{\phi_1}{\mathfrak{f}_1}  + \frac{q_{22}}{N_2}\,\frac{\phi_2}{\mathfrak{f}_2} \right)\right]
\label{Vlate3new}
\ee
where
\be
\mathfrak{f}_1\equiv \frac{f_1}{N_1} = \frac{1}{2\sqrt{2}\pi\langle\tau_1\rangle}\qquad\text{and}\qquad \mathfrak{f}_2\equiv \frac{f_2}{N_2} = \frac{1}{2\pi\langle\tau_2\rangle}
\ee
This potential has a flat direction in the limit of perfect alignment \cite{hep-ph/0409138}, when $\det q=q_{11}q_{22}-q_{12} q_{21}=0$. This can be easily seen by performing the following field redefinition
\begin{equation}
\left(\begin{matrix} 
\phi_L \\ 
\phi_H 
\end{matrix}\right) = \frac{1}{\sqrt{\frac{q_{21}^2}{\mathfrak{f}_1^2} + \frac{q_{22}^2}{\mathfrak{f}_2^2}}}
\left(\begin{matrix} 
s \frac{q_{22}}{\mathfrak{f}_2} & -s\frac{q_{21} }{\mathfrak{f}_1}  \\ 
\frac{ q_{21} }{\mathfrak{f}_1} & \frac{q_{22}}{ \mathfrak{f}_2} 
\end{matrix}\right)
\left(\begin{matrix} 
\phi_1 \\ 
\phi_2 
\end{matrix}\right)
\end{equation}
where $s=\pm 1$ coincides with the sign of $\det q$.  This change of basis brings (\ref{Vlate3}) into the simplified form
\be
V_\text{late}\simeq  \hat{\Lambda}_2^4\left[1-\cos \left(q_{2H} \,\frac{\phi_H}{\mathfrak{f}_H}\right)\right]+\hat{\Lambda}_1^4\left[1-\cos \left(q_{1H} \,\frac{\phi_H}{\mathfrak{f}_H}  + q_L\,\frac{\phi_L}{\mathfrak{f}_L} \right)\right]
\label{Vlate4new}
\ee
where
\begin{eqnarray}
 q_{1H} &\equiv& \frac{N_2}{N_1}\left(\frac{q_{11}q_{21}/\mathfrak{f}_1^2+q_{12}q_{22}/\mathfrak{f}_2^2}{q_{21}^2 /\mathfrak{f}_1^2 +q_{22}^2 /\mathfrak{f}_2^2}\right) q_{2H} \qquad q_{2H}\equiv \frac{1}{N_2}\sqrt{q_{21}^2 + q_{22}^2}    \nonumber \\
q_L &\equiv& \frac{|\det q |}{N_1 \sqrt{q_{21}^2 + q_{22}^2}}
\qquad \mathfrak{f}_H \equiv  \sqrt{\frac{ q_{21}^2 + q_{22}^2}{q_{21}^2 /\mathfrak{f}_1^2+ q_{22}^2 /\mathfrak{f}_2^2 }}\qquad \mathfrak{f}_L\equiv \frac{\mathfrak{f}_1 \mathfrak{f}_2}{\mathfrak{f}_H}
\end{eqnarray}
In the alignment limit we have  $\det q=0$,  and so $q_L = 0$ and the potential (\ref{Vlate4new}) depends only  on $\phi_H$, showing that $\phi_L$ becomes an exactly flat direction. The presence of a massless axion can also be seen from the fact that the determinant of the mass-squared matrix vanishes in the alignment limit
\begin{equation}
\left.\det\left(\begin{matrix} 
\frac{\partial^2 V_{\rm late}}{\partial\phi_H^2} & \frac{\partial^2 V_{\rm late}}{\partial\phi_H\partial\phi_L}  \\ 
\frac{\partial^2 V_{\rm late}}{\partial\phi_H\partial\phi_L} & \frac{\partial^2 V_{\rm late}}{\partial\phi_L^2} 
\end{matrix}\right)\right|_{\rm min} = \left(\hat{\Lambda}_1\hat{\Lambda}_2\right)^4\left(\frac{q_{1H} q_L}{\mathfrak{f}_H \mathfrak{f}_L}\right)^2 = 0 \qquad\text{for}\qquad q_L=0\,.  
\end{equation}
To achieve slow-roll at the correct scale, quintessence should be a {\it nearly} flat direction of the potential, corresponding to approximate alignment. In order for this to happen, we assume that the corresponding charges can be chosen such that $q_L \ll 1$. Note that also a relatively large $N_1$ helps to achieve a small $q_L$. The axion $\phi_H$ becomes hierarchically heavier than $\phi_L$ for $q_L\ll 1$. 

Quintessence can now be realised around the saddle point where the $\phi_H$ direction is stable and $\phi_L$ is close to the maximum of its effective potential. Assuming for definiteness that $\hat{\Lambda}_2 \gg \hat{\Lambda}_1$, the mass hierarchy around this saddle point is
\be
m_H\simeq q_{2H}\,\frac{\hat{\Lambda}_2^2}{\mathfrak{f}_H}\gg |m_L| \simeq q_L\, \frac{\hat{\Lambda}_1^2}{\mathfrak{f}_L}\,.
\label{maxionsnew}
\ee
Due to this hierarchy, the leading order potential depends just on $\phi_H$ and goes as
\begin{equation}
V(\phi_H)\simeq \hat{\Lambda}_2^4\left[1-\cos\left(\frac{\phi_H}{f_H}\right)\right]
\label{Veffhnew}
\end{equation}
where the effective decay constant and mass of $\phi_H$ are given by
\begin{equation}
f_H\equiv \frac{\mathfrak{f}_H}{q_{2H}} = \frac{N_2 }{\sqrt{q_{21}^2 /\mathfrak{f}_1^2+ q_{22}^2 /\mathfrak{f}_2^2}} \qquad\text{and}\qquad m_H \simeq \frac{\hat{\Lambda}_2^2}{f_H}
\label{fhnew}
\end{equation}
The heavy mode $\phi_H$ can therefore be integrated out by setting it to its minimum at $\langle\phi_H\rangle = 0$. The effective potential for the light axion $\phi_L$ then becomes
\begin{equation}
V(\phi_L)\simeq \hat{\Lambda}_1^4\left[1-\cos\left(\frac{\phi_L}{f_L}\right)\right]
\label{Veffnew}
\end{equation}
where
\begin{equation}
f_L\equiv \frac{\mathfrak{f}_L}{q_L} = \frac{N_1 \mathfrak{f}_1 \mathfrak{f}_2}{|\det q |} \sqrt{q_{21}^2 /\mathfrak{f}_1^2+ q_{22}^2 /\mathfrak{f}_2^2}
\label{flnew}
\end{equation}
We see how the approximate alignment gives rise to a large decay constant, $f_L\gg 1$, via the KNP proposal \cite{hep-ph/0409138}. In Sec. \ref{sec:constraintsnew} these expressions for the decay constants will allow us to easily compare with observational constraints on ultra-light axions.

In this scenario, the heavy axion $\phi_L$ is stable and oscillates about equilibrium point, providing a candidate for (a fraction of) dark matter today. On the other hand, the light axion $\phi_L$ is displaced from its minimum and plays the role of the quintessence field. The corresponding slow-roll parameters go as $\epsilon \simeq |\eta| \simeq 1/f_L^2$. Note that $\eta$ is small in absolute value, and so the light axion can be identified with a quintessence field in slow-roll.

\subsection{Phenomenological and theoretical constraints}
\label{sec:constraintsnew}

Let us now impose a number of phenomenological and theoretical constraints on our model in order to find a consistent set of parameters and to extract any important predictions.

\subsubsection*{Dark matter and dark energy constraints}
\label{DMsecnew}

The late universe physics predicts a dark matter candidate via the heavier of the two axions. Let us assume that this dark matter field starts oscillating before matter-radiation equality, which occurs at a Hubble scale of order $H_\text{eq} \simeq 10^{-54}\,M_p\simeq 10^{-27}$ eV. This imposes the lower bound $m_H \gtrsim H_\text{eq}$. The abundance of axion dark matter today is given by \cite{Marsh:2015xka}
\be 
\Omega_H=\frac16 (9 \Omega_r)^{3/4} \sqrt{\frac{m_H}{10^{-60} M_p}}\, \left(\frac{\delta \phi_H}{M_p}\right)^2
\label{abundancenew}
\ee
where $\delta \phi_H$ parametrises the initial displacement from the minimum of the heavy axion. Using the known abundances of radiation $\Omega_r \simeq 0.8 \times 10^{-4}$ and dark matter $\Omega_m\simeq 0.26$ \cite{Planck:2018jri}, we can derive the following expressions for the proportion of dark matter coming from the heavy axion
\begin{eqnarray}
\frac{\Omega_H}{\Omega_m}  &\simeq& 7.14 \times 10^{-5} \left(\frac{H_{\rm inf}}{M_p}\right)^2 \sqrt{\frac{m_H}{10^{-60} M_p}} \left(\frac{2\pi\delta \phi_H}{H_\text{inf}}\right)^2 \nonumber \\ 
&\simeq& 2.78 \times 10^{-6}   \left(\frac{f_H}{0.01 M_p}\right)^2 \sqrt{\frac{m_H}{10^{-60} M_p}} \left(\frac{\delta \phi_H}{\pi f_H}\right)^2 \ .
\label{ax-fractionnew}
\end{eqnarray}
Note that we expect $\frac{H_\text{inf}}{2\pi} \lesssim \delta \phi_H \lesssim  \pi f_H$, with the lower bound coming from quantum diffusion during inflation \cite{Hardwick:2017fjo,Cicoli:2021skd}. This translates into the following inequalities for the axion dark matter fraction:
\be 
5.85\times 10^{-15} \sqrt{\frac{m_H}{10^{-60} M_p}}  \left(\frac{H_{\rm inf}}{ 0.905 \times  10^{-5} M_p}\right)^2 \lesssim \frac{\Omega_H}{\Omega_m} \lesssim 2.78 \times 10^{-6} \sqrt{\frac{m_H}{10^{-60} M_p}} \left(\frac{f_H}{0.01 M_p}\right)^2
\label{ax-fractionineqnew}
\ee
where we have normalised  $H_\text{inf}$ relative to the inflationary scale $ 0.905 \times  10^{-5}\,M_p$  appearing  in \eqref{Hinf}.

\begin{figure}[ht]
\begin{center}
\includegraphics[width=0.99\textwidth]{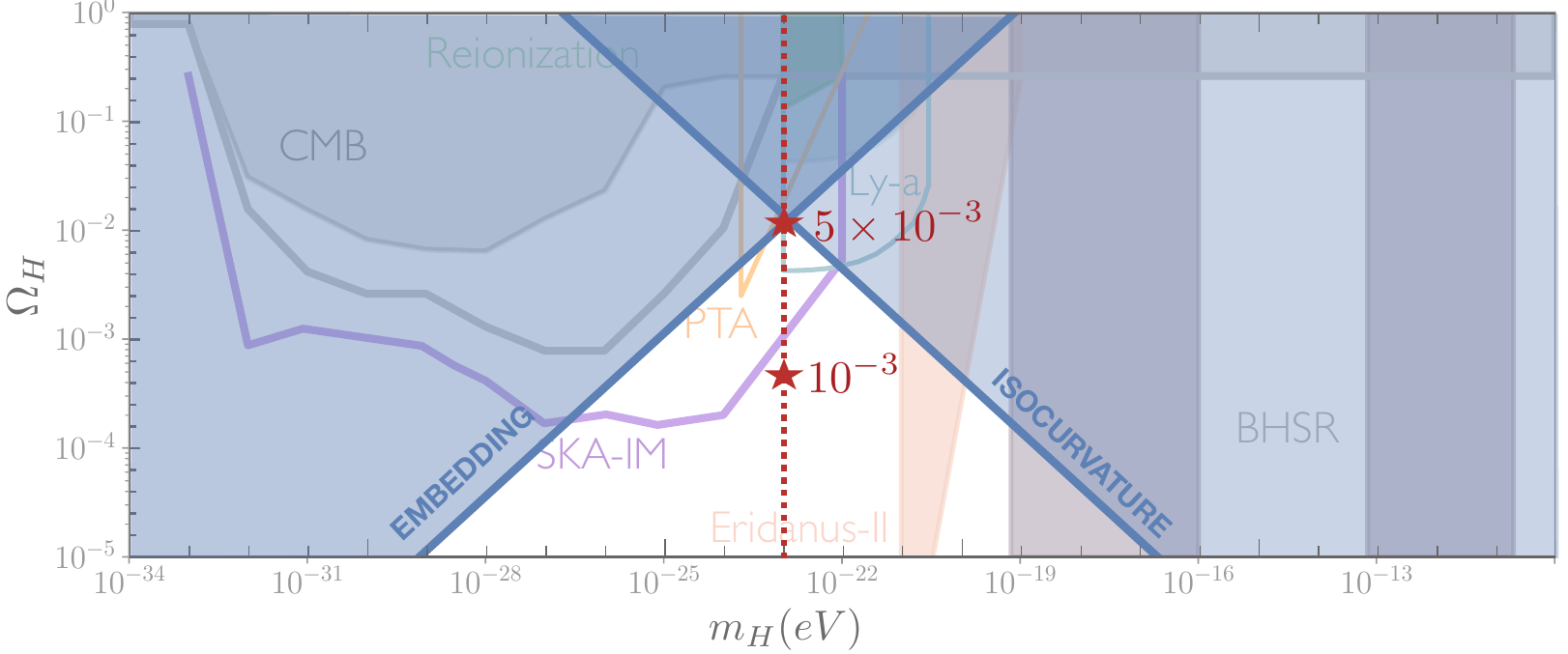}
\caption{Cosmological and astrophysical bounds on the axion dark matter candidate $\phi_H$. The blue shaded areas are excluded by the isocurvature bound (right) and the embedding constraint (left) forcing $f_H<0.005\,M_p$ for the present model where $H_{\rm inf}\simeq 10^{-5} M_p$. The dashed vertical line corresponds to $m_H=10^{-23}$ eV, for different values of $f_H$ indicated by the stars, assuming $\delta \phi_H= \pi f_H$. The figure is adapted from \cite{Marsh:2021jmi}.}
\label{fig:DMconstraints}
\end{center}
\end{figure}

The abundance of axion dark matter is also subject to isocurvature constraints if the Peccei-Quinn symmetry for the heavy axion is broken during inflation, $f_H \gtrsim H_\text{inf}$. If this is the case, the axion will generate isocurvature perturbations with amplitude \cite{Marsh:2015xka}
\be
\label{AInew}
\mathcal{A}_I =\left(\frac{\Omega_H}{\Omega_m}\right)^2 \left(\frac{H_\text{inf}}{\pi \delta \phi_H}\right)^2,
\ee
that is bound to satisfy $\mathcal{A}_I/\mathcal{A}_s<0.038$ \cite{Marsh:2015xka}.  If we use \eqref{AInew} to express $\delta \phi_H^2$ in terms of the  amplitude for isocurvature perturbations,  and substitute the result into the formula for the axion abundance relation \eqref{abundancenew}, we can derive another  upper bound on the fraction of dark matter coming from the heavy axion:
\be 
\frac{\Omega_H}{\Omega_m} \lesssim 3395 \sqrt{\frac{10^{-60} M_p}{m_H}}\left(\frac{0.0005}{
\epsilon}\right) \ .
\label{upperboundnew}
\ee
When compared with the left hand inequality \eqref{ax-fractionineqnew}, this imposes a relatively weak upper bound on the mass of the heavy axion.  We can say much more by adding the two bounds \eqref{ax-fractionineqnew} and \eqref{upperboundnew} to other dark matter constraints on axions, as presented in Fig. \ref{fig:DMconstraints}, remembering that we have also assumed that $m_H>H_\text{eq} \simeq 10^{-27}$ eV in order that the heavy axion becomes dynamical before matter-radiation equality.

If we now wish to maximise the fraction of axion dark matter, the isocurvature constraints in Fig. \ref{fig:DMconstraints} seem to favour a heavy axion with a mass of around  $10^{-25}$ eV which would correspond to $f_H\simeq 0.1\,M_p$ for $\delta \phi_H\simeq \pi f_H$. Indeed, for $m_H \simeq  10^{-25}$ eV and $f_H \simeq  0.075\,M_p$,  we see from \eqref{ax-fractionnew} that the heavy axion could, in principle, be all of dark matter. However, in the context of the current model, such conclusions are too quick. As we will show in Sec. \ref{sec:examplesnew}, an additional constraint arises from requiring a consistent embedding in a string compactification. In fact, imposing that the decay constant $f_1$ is real sets a strong upper bound on  $f_H\lesssim 0.005\, M_p$ which, in turn, implies that the heavy axion can be at most around $0.1\%$ of dark matter for $m_H \simeq 10^{-25}$ eV. Due to this embedding constraint, combined with the isocurvature one, the heavy axion $\phi_H$ can at most be $5\%$ of dark matter for $m_H\simeq 10^{-23}$ eV, as we will show in Sec. \ref{sec:examplesnew}.

Let us now focus on dark energy constraints. The cosmological constant scale is set by the size of the late time potential at the corresponding saddle, giving
\be
\hat{\Lambda}_1 \sim 10^{-30}\,M_p\,.
\ee
As already pointed out, axion hilltop quintessence can be ruined by quantum diffusion during inflation unless $\Delta_{\max}(f_L)>H_\text{inf} \simeq 10^{-5} M_p$. As can be seen from Fig. \ref{fig:f_vs_Phi0} (where now $f$ should be identified with $f_L$), this implies that for the present model $f_L\gtrsim 0.08\, M_p$.

\subsubsection*{UV embedding constraints and explicit examples}
\label{sec:examplesnew}

In this section we combine early and late time constraints on the model with the goal of finding explicit examples that allow for observationally viable cosmology. The scales of dark matter and dark energy are determined by the magnitudes of the axionic potential \eqref{LamInew} which can be rewritten as  
\begin{equation}
\label{LInew}
\hat{\Lambda}_i^4=\frac{4 |W_0| A_i }{\mathcal{V}^2}\, S_i\, e^{-S_i}\qquad\text{with}\qquad S_i\equiv \frac{1}{N_i}\left(\frac{q_{i1}}{\sqrt{2}\,\mathfrak{f}_1}+\frac{q_{i2}}{\mathfrak{f}_2}\right).
\end{equation}
Note that small differences in the arguments of the exponentials can lead to large hierarchies, as necessary for describing dark matter and dark energy. Recall that we assume that $\hat{\Lambda}_2\gg \hat{\Lambda}_1$.

As we have seen in Sec. \ref{InflConstr}, for compactifications with natural $\mathcal{O}(1)$ values of the microscopic parameters, getting the correct inflationary scale requires $\mathcal{V}\simeq 10^3$. If one further assumes that the compactification is isotropic at the minimum after the end of inflation, $\langle\tau_1\rangle\sim \langle\tau_2\rangle$, one is led to the conclusion that $\langle\tau_1\rangle\sim \langle\tau_2\rangle\simeq 100$. For concreteness, in what follows we set $\mathcal{V}=10^3$ and allow $\langle\tau_1\rangle\in [80,100]$. Let us stress that, for $\langle\tau_1\rangle\sim \langle\tau_2\rangle$, the inflationary condition (\ref{Cond4}) can only  be met  for $\Pi_2=0$. This corresponds to the case $\mathcal{R}=0$ in the inflationary potential \eqref{Vinf3}. The corresponding scale of inflation and slow-roll parameter for $N_e\simeq 52$ efoldings are respectively $H_\text{inf} \simeq 0.86 \times 10^{-5} M_p$ and $\epsilon \simeq 0.0004$.

The decay constants for the canonically normalised axions $\phi_H$ and $\phi_L$ are given by \eqref{fhnew} and \eqref{flnew}. 
Let us define $\delta \equiv 100 f_H/M_p$ such that $\delta=1$ corresponds to a GUT scale decay constant, $f_H=0.01\,M_p$. This immediately imposes the following constraint
\begin{equation}
\label{fHconnew}
    \frac{q_{21}^2}{\mathfrak{f}_1^2} +\frac{q_{22}^2}{\mathfrak{f}_2^2}  =\left( 100\,\frac{N_2}{\delta}\right)^2
\end{equation}
and so generically we expect each term to scale as $100 N_2/\delta$. Requiring $f_L\ge 0.1\,M_p$ imposes another constraint
\be
\frac{|\det q |}{\mathfrak{f}_1\mathfrak{f}_2} \lesssim \frac{10^3}{\delta} N_1 N_2\,.
\label{eq:constraintKNPnew}
\ee
In the absence of cancellations, this would also be of order $10^4 N_1 N_2/\delta^2$, indicating that we need a tuning of order $\delta/10$. A higher level of tuning will be required (and can be accommodated in the present model) if the typical size of $f_H$ is smaller than the GUT scale, $\delta\ll1$. To be more precise, we can relate each combination $S_i$ to $\hat{\Lambda}_i$ according to \eqref{LInew}. Dark energy constraints impose $\hat{\Lambda}_1 \sim 10^{-30}\,M_p$, and so assuming $|W_0|\sim A_1\sim A_2 \sim \mathcal{O}(1)$ and $\mathcal{V} \simeq 10^3$, we find 
\begin{equation} 
\frac{q_{11}}{\sqrt{2} \mathfrak{f}_1}+\frac{q_{12}}{\mathfrak{f}_2} \simeq 269 \,N_1\,.
\label{eq:constraintLambdaDEnew}
\end{equation}
We now consider the role of the heavy axion, normalising its mass relative to the scale $10^{-25}$ eV. To this end, we introduce $\mu= m_H/(10^{-25} \text{eV})$ and note that 
\be
\hat{\Lambda}_2^4 \simeq m_H^2 f_H^2= 1.69\times 10^{-109} \left(\mu\delta\right)^2
\ee
Once again assuming $|W_0|\sim A_1\sim A_2 \sim \mathcal{O}(1)$ and $\mathcal{V} \simeq 10^3$, we now find
\begin{equation} 
\frac{q_{21}}{\sqrt{2} \mathfrak{f}_1}+\frac{q_{22}}{\mathfrak{f}_2}
\simeq \left[244 - 2\ln(\mu\delta)\right] N_2\,.
\label{eq:constraintLambdaDMnew}
\end{equation}
This relation, together with \eqref{fHconnew}, can fix $q_{21}/\mathfrak{f}_1$ and $q_{22}/\mathfrak{f}_2$. Indeed, we can eliminate $q_{22}/\mathfrak{f}_2$ in \eqref{fHconnew} to derive the following quadratic equation for $q_{21}/\mathfrak{f}_1$
\begin{equation}
\frac32 \left(\frac{q_{21}}{\mathfrak{f}_1}\right)^2 -\sqrt{2}N_2(244-2 
\ln(\mu \delta))\left(\frac{q_{21}}{\mathfrak{f}_1}\right)+\left[(244-2 
\ln(\mu \delta))^2-\frac{10^4}{\delta^2}\right] N_2^2=0
\end{equation}
Of course, $f_1$ is a real number, and so this equation must have real roots. This allows us to derive an important constraint on the decay constant $f_H$. In particular we find
\begin{equation}
    \delta\left(1-0.82 \times 10^{-3} \ln (\mu \delta)\right) \le  0.50
\end{equation}
which amounts to $f_H \le 0.005\,M_p$ for a very large range of masses for the heavy axion. Thanks to the inequality in \eqref{ax-fractionineqnew}, this translates into a strict upper limit on the fraction of dark matter:
\begin{equation}
\label{newfracboundnew}
    \frac{\Omega_H}{\Omega_m} \lesssim 6.95 \times 10^{-7} \sqrt{\frac{m_H}{10^{-60}\,M_p}} 
\simeq 0.045\,  \sqrt{\frac{m_H}{10^{-23} \,\text{eV}}} \ .
\end{equation}
Even though the isocurvature constraints of Fig. \ref{fig:DMconstraints} suggest an optimal case of $m_H \simeq 10^{-25}$ eV, the microscopic details of this particular model mean that an axion this heavy can only account for at most $0.45\%$ of dark matter.  We can get a larger fraction by making the axion heavier, reaching up to about $4.5\%$ of dark matter for $m_H \simeq 10^{-23}$ eV.

\begin{figure}
\centering
\includegraphics[width=.39\linewidth]{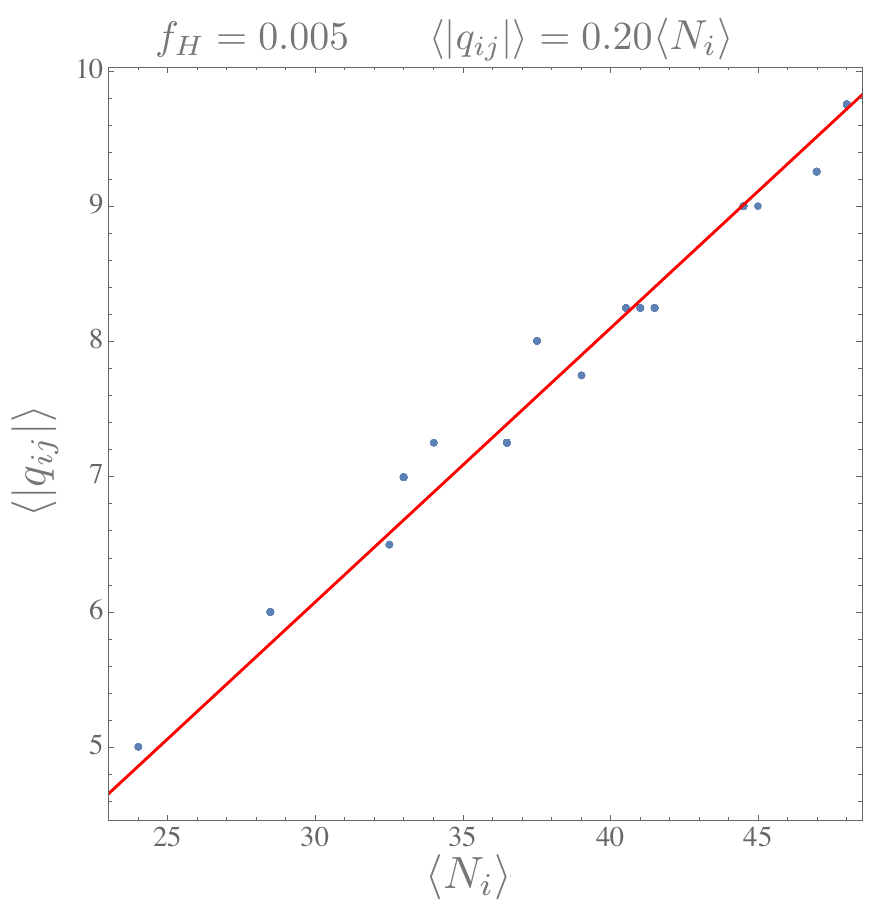}
\includegraphics[width=.4\linewidth]{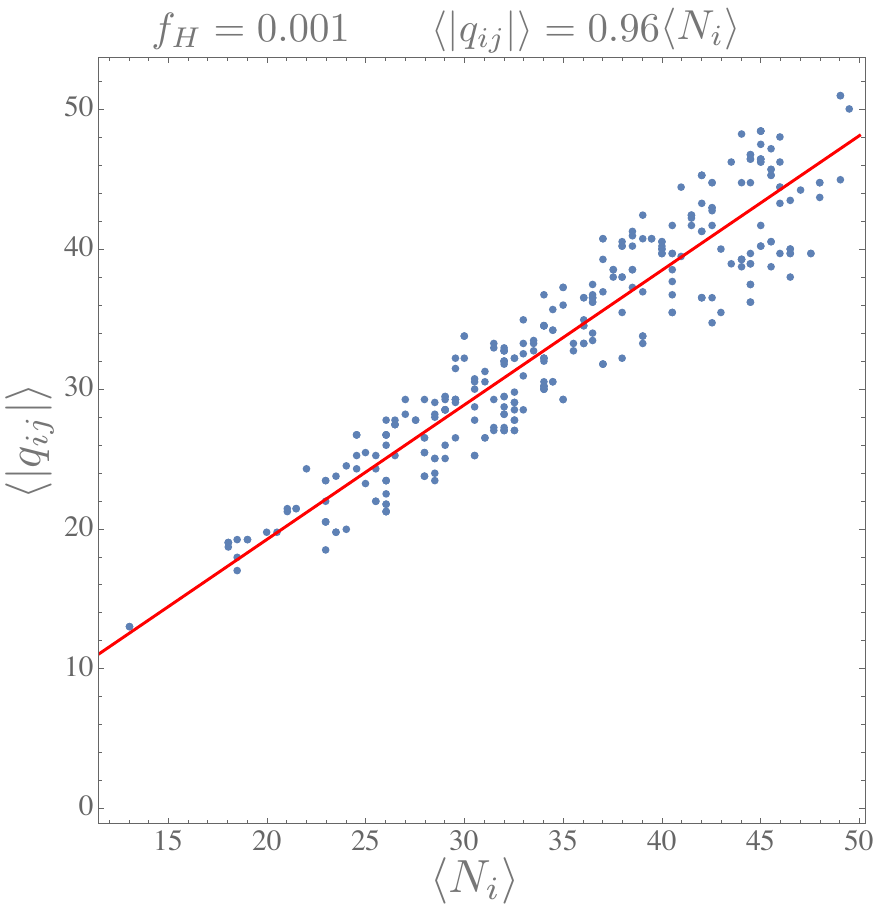}
\includegraphics[width=.39\linewidth]{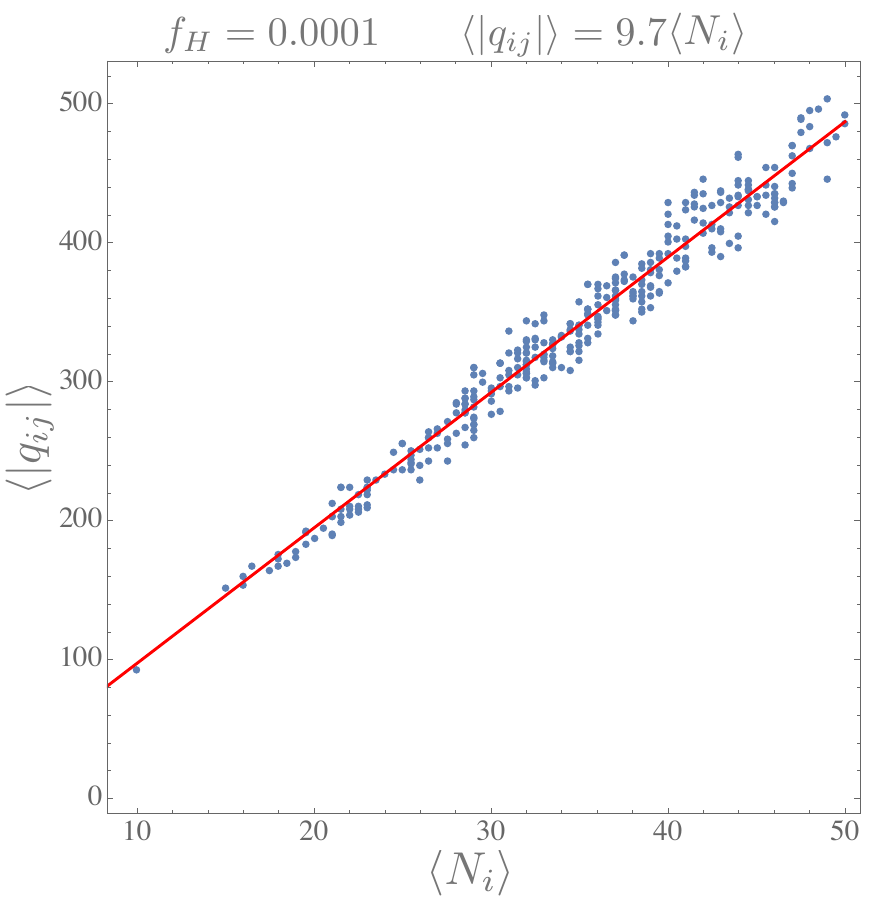}
\caption{Mean gauge group rank $\langle N_i\rangle$ and mean modulus of the winding numbers $\langle|q_{ij}|\rangle$ for $m_H= 10^{-23}$ eV and $f_L\simeq 0.1\,M_p$, of which the examples in Tab. \ref{tab:NumEx} are particular examples. The red lines correspond to the best linear fits.}
\label{fig:stats1}
\end{figure}

So far, we have only used the fact that $\mathfrak{f}_1$ is real. To arrive at a genuine microscopic model giving rise to the desired late time cosmology, we also need to match $f_H$ and $f_L$ to the underlying microscopic parameters. Recall that we have integer winding numbers, $q_{ij} \in \mathbb{Z}$, while $a_i=2\pi$ for a D3-instanton while $a_i=2\pi/N_i$ for gaugino condensation on a stack of $N_i$ D7-branes. 
In Tab. \ref{tab:NumEx} we present three numerical examples for distinct choices of $f_H$ for $m_H\simeq 10^{-23}$ eV. In Fig. \ref{fig:stats1} we plot the mean gauge group rank $\langle N_i \rangle$ and the mean modulus of the winding numbers $\langle |q_{ij}| \rangle$ corresponding to the choices of $m_H$ and $f_H$ in Tab. \ref{tab:NumEx} in order to demonstrate that multiple solutions exist and that the values in the table are not special in any way.

\begin{table}[ht]
\centering
\resizebox{!}{2.6cm}{
\begin{tabular}{c|c|c|c|c||c|c|c}
$\langle\tau_1\rangle$ & $\langle\tau_2\rangle$& $N_1$ & $N_2$ & $q_{ij}$ & $f_L$ & $f_H$ & $\Omega_H^{\max}/\Omega_m$\\
\hline
\hline
106.13 & 97.07 & 23& 42&  $\left(\begin{array}{cc}2 & 8 \\3 & 13\\\end{array}\right)$ & 0.17& 0.005& 0.045 \\
\hline
95.08& 102.56& 20& 16& $\left(\begin{array}{cc}  -19& 26 \\-13 & 18 \\\end{array}\right)$ & 0.15& 0.001 &0.002 \\
\hline
94.13& 103.07& 11& 24 & $\left(\begin{array}{cc}-122 & 116\\ -231& 220\\ \end{array}\right)$ &0.11& 0.0001 & $2\times 10^{-5}$ \\
\hline
\end{tabular}}
\caption{Numerical examples for $m_H \simeq 10^{-23}$ eV. The decay constants $f_H$ and $f_L$ are given in Planck units. $\Omega_H^{\max}$ is the heavy axion contribution to the dark matter density, assuming maximal misalignment, $\delta \phi_H = \pi f_H/M_p$, as per Eq. \eqref{ax-fractionineqnew}.}
\label{tab:NumEx}
\end{table}

\subsubsection*{Compatibility with the axion weak gravity conjecture}

The original weak gravity conjecture \cite{Arkani-Hamed:2006emk} states that a $U(1)$ gauge theory can only be embedded in a consistent quantum theory of gravity if there is a particle whose charge to mass ratio $q/m>1$ in Planck units. The claim is justified in order to ensure the decay of extremal black holes, avoiding remnants, and can be generalised to multiple $U(1)$s via the convex hull condition \cite{Cheung:2014vva}. To express it, we consider the $i$-th particle with mass $m_i$ and charge vector $\vec{q}_i$, and  define the charge-to-mass ratio as $\vec{z}_i=\vec{q}_i/m_i$. The weak gravity conjecture now requires that the convex hull spanned by the set $\pm \vec{z}_i$ should contain the unit ball. 

In \cite{Arkani-Hamed:2006emk}, the  weak gravity conjecture is generalised to $p$-form gauge fields, with a  charged $(p-1)$-brane generalising the charged particle and its tension generalising the mass. The idea can be extended to axions which are $0$-forms, with the corresponding $(-1)$-branes identified with instantons with 
`mass' given by the instanton action. For a single axion with quantised charge $q$, decay constant $f$, and instanton action $S$, the analogue of a charge-to-mass ratio can be defined as $z\equiv q/(f\,S)$. According to the axion weak gravity conjecture, this ratio should be bounded below  by the charge-to-mass ratio of an extremal gravitational instanton, assumed here to be order one\footnote{There is some confusion over the choice of extremal instanton solution. A discussion can be found in \cite{Harlow:2022gzl}.}. As is well known, for the instanton action to remain suppressed, this implies a sub-Planckian decay constant.   

For multiple axions, as in our model, the situation is inevitably more complex, just as it was for the $U(1)$s. We have quantised charges $q_{ij}$, decay constants $\mathfrak{f}_1$ and $\mathfrak{f}_2$ and instanton actions $S_i = a_i q_{ij} \tau_j$ with $i=1,2,3$. We can then define the $i$-th charge-to-mass ratio vector, $z_i$, as: 
\be 
\vec{z}_i= \frac{1}{S_i}\left( \frac{q_{i1}}{\mathfrak{f}_1}, \frac{q_{i2}}{\mathfrak{f}_2}\right) \qquad i=1,2,3\,.
\label{z}
\ee
The axion weak gravity conjecture is now expressed in terms of the convex hull condition \cite{Rudelius:2015xta}: the convex hull spanned by the set $\pm \vec z_i$ should contain the ball of radius $r_\text{CHC}$, where $r_\text{CHC}$ is the norm of the charge-to-mass vector for the extremal instanton, assumed here to be of order one.
By themselves, two aligned axions cannot satisfy this condition since this corresponds to the limit where the charge-to-mass ratios are parallel vectors $\vec{z_1} \propto \vec{z_2}$. This makes the convex hull spanned by $\pm \vec z_1$ and $\pm \vec z_2$ line-like and therefore unable to contain the ball of radius $r_\text{CHC}$ \cite{Rudelius:2015xta, 2106.09853}. The solution relies on the presence of a third instanton with action $S_3$ whose charge-to-mass vector $\vec z_3$ points in an orthogonal direction. This will have no impact on the late time cosmological dynamics described in Sec. \ref{sec:late} as long as $S_3\gg S_1,S_2$ which can be easily achieved if $\vec z_3$ is in the first/third quadrant of the Euclidean plane \cite{2106.09853}. By construction, $\vec z_1$ and $\vec z_2$ are aligned and almost orthogonal to $\vec z_3$, and so must lie in the second/fourth quadrant, resulting in
\be
\frac{|\vec{z}_i \cdot \vec{z}_3|}{|\vec{z}_i || \vec{z}_3|} \ll 1\qquad \forall i=1,2\,.
\ee
In order to ensure that the convex hull contains the ball of radius $r_\text{CHC}$, we require that $|\vec z_i| >r_\text{CHC}$ for $i=3$ and at least one of $i=1,2$. As an illustrative example let us consider the second line of Tab. \ref{tab:NumEx}. For this case one can show that 
\begin{equation}
\vec{z}_1=(-59.41, 62.01)\qquad\text{and}\qquad  \vec{z}_2= (-45.85, 48.42) \, .
\end{equation}
It is straightforward to check that the angle between these two vectors is $\simeq 0.0059\ll 1$, as it should for KNP alignment, and that the convex hull is very narrow and does not fully contain the unit ball. In order to satisfy the convex hull condition, the vector $\vec{z}_3$ has to be orthogonal to $\vec{z_1}$ and has to have $|\vec{z}_3|\ge r_\text{CHC}=\sqrt{3/2}$. This can be achieved with sufficient precision by choosing $N_3=2$, $q_{31}=7$ and $q_{32}=8$. The instanton actions for this example are $S_1=270$, $S_2=240$ (in accordance with \eqref{eq:constraintLambdaDEnew} and \eqref{eq:constraintLambdaDMnew}) and $S_3=4669$, giving
\begin{equation}
\vec{z}_3=(1.267, 1.104)\qquad \frac{|\vec{z}_2 \cdot \vec{z}_3|}{|\vec{z}_2 || \vec{z}_3|}\simeq 0.04 \qquad |\vec{z}_3|\simeq 1.68 > r_\text{CHC}=\sqrt{3/2}\,.
\end{equation}

We note that for the largest values of $f_H$, corresponding to the first line of Tab. \ref{tab:NumEx}, all the entries of the charge matrix are positive, implying that $z_1$ and $z_2$ would lie in the first/third quadrant. In such cases, satisfying the convex hull condition with an instanton of large action becomes more challenging.

\bibliography{main}

\providecommand{\href}[2]{#2}\begingroup\raggedright\begin{thebibliography}{100}

\bibitem{CMB}
{\scshape Planck} collaboration, \emph{{Planck 2018 results. VI. Cosmological parameters}}, \href{https://doi.org/10.1051/0004-6361/201833910}{\emph{Astron. Astrophys.} {\bfseries 641} (2020) A6} [\href{https://arxiv.org/abs/1807.06209}{{\ttfamily 1807.06209}}].

\bibitem{Sn1}
{\scshape Supernova Cosmology Project} collaboration, \emph{{Measurements of $\Omega$ and $\Lambda$ from 42 high redshift supernovae}}, \href{https://doi.org/10.1086/307221}{\emph{Astrophys. J.} {\bfseries 517} (1999) 565} [\href{https://arxiv.org/abs/astro-ph/9812133}{{\ttfamily astro-ph/9812133}}].

\bibitem{Sn2}
{\scshape Supernova Search Team} collaboration, \emph{{Observational evidence from supernovae for an accelerating universe and a cosmological constant}}, \href{https://doi.org/10.1086/300499}{\emph{Astron. J.} {\bfseries 116} (1998) 1009} [\href{https://arxiv.org/abs/astro-ph/9805201}{{\ttfamily astro-ph/9805201}}].

\bibitem{daniel}
D.~Baumann, \emph{{Inflation}},  in \emph{{Theoretical Advanced Study Institute in Elementary Particle Physics}: {Physics of the Large and the Small}}, pp.~523--686, 2011, \href{https://doi.org/10.1142/9789814327183_0010}{DOI} [\href{https://arxiv.org/abs/0907.5424}{{\ttfamily 0907.5424}}].

\bibitem{ed}
E.J.~Copeland, M.~Sami and S.~Tsujikawa, \emph{{Dynamics of dark energy}}, \href{https://doi.org/10.1142/S021827180600942X}{\emph{Int. J. Mod. Phys. D} {\bfseries 15} (2006) 1753} [\href{https://arxiv.org/abs/hep-th/0603057}{{\ttfamily hep-th/0603057}}].

\bibitem{quin}
S.~Tsujikawa, \emph{{Quintessence: A Review}}, \href{https://doi.org/10.1088/0264-9381/30/21/214003}{\emph{Class. Quant. Grav.} {\bfseries 30} (2013) 214003} [\href{https://arxiv.org/abs/1304.1961}{{\ttfamily 1304.1961}}].

\bibitem{cliff}
C.P.~Burgess, \emph{{The Cosmological Constant Problem: Why it's hard to get Dark Energy from Micro-physics}},  in \emph{{100e Ecole d'Ete de Physique: Post-Planck Cosmology}}, pp.~149--197, 2015, \href{https://doi.org/10.1093/acprof:oso/9780198728856.003.0004}{DOI} [\href{https://arxiv.org/abs/1309.4133}{{\ttfamily 1309.4133}}].

\bibitem{tony}
A.~Padilla, \emph{{Lectures on the Cosmological Constant Problem}},  \href{https://arxiv.org/abs/1502.05296}{{\ttfamily 1502.05296}}.

\bibitem{DM1}
J.L.~Feng, \emph{{Dark Matter Candidates from Particle Physics and Methods of Detection}}, \href{https://doi.org/10.1146/annurev-astro-082708-101659}{\emph{Ann. Rev. Astron. Astrophys.} {\bfseries 48} (2010) 495} [\href{https://arxiv.org/abs/1003.0904}{{\ttfamily 1003.0904}}].

\bibitem{DM2}
G.~Bertone and D.~Hooper, \emph{{History of dark matter}}, \href{https://doi.org/10.1103/RevModPhys.90.045002}{\emph{Rev. Mod. Phys.} {\bfseries 90} (2018) 045002} [\href{https://arxiv.org/abs/1605.04909}{{\ttfamily 1605.04909}}].

\bibitem{Cicoli:2023opf}
M.~Cicoli, J.P.~Conlon, A.~Maharana, S.~Parameswaran, F.~Quevedo and I.~Zavala, \emph{{String cosmology: From the early universe to today}}, \href{https://doi.org/10.1016/j.physrep.2024.01.002}{\emph{Phys. Rept.} {\bfseries 1059} (2024) 1} [\href{https://arxiv.org/abs/2303.04819}{{\ttfamily 2303.04819}}].

\bibitem{kklt}
S.~Kachru, R.~Kallosh, A.D.~Linde and S.P.~Trivedi, \emph{{De Sitter vacua in string theory}}, \href{https://doi.org/10.1103/PhysRevD.68.046005}{\emph{Phys. Rev. D} {\bfseries 68} (2003) 046005} [\href{https://arxiv.org/abs/hep-th/0301240}{{\ttfamily hep-th/0301240}}].

\bibitem{Aparicio:2015psl}
L.~Aparicio, F.~Quevedo and R.~Valandro, \emph{{Moduli Stabilisation with Nilpotent Goldstino: Vacuum Structure and SUSY Breaking}}, \href{https://doi.org/10.1007/JHEP03(2016)036}{\emph{JHEP} {\bfseries 03} (2016) 036} [\href{https://arxiv.org/abs/1511.08105}{{\ttfamily 1511.08105}}].

\bibitem{Crino:2020qwk}
C.~Crin\`o, F.~Quevedo and R.~Valandro, \emph{{On de Sitter String Vacua from Anti-D3-Branes in the Large Volume Scenario}}, \href{https://doi.org/10.1007/JHEP03(2021)258}{\emph{JHEP} {\bfseries 03} (2021) 258} [\href{https://arxiv.org/abs/2010.15903}{{\ttfamily 2010.15903}}].

\bibitem{Burgess:2003ic}
C.P.~Burgess, R.~Kallosh and F.~Quevedo, \emph{{De Sitter string vacua from supersymmetric D terms}}, \href{https://doi.org/10.1088/1126-6708/2003/10/056}{\emph{JHEP} {\bfseries 10} (2003) 056} [\href{https://arxiv.org/abs/hep-th/0309187}{{\ttfamily hep-th/0309187}}].

\bibitem{Braun:2015pza}
A.P.~Braun, M.~Rummel, Y.~Sumitomo and R.~Valandro, \emph{{De Sitter vacua from a D-term generated racetrack potential in hypersurface Calabi-Yau compactifications}}, \href{https://doi.org/10.1007/JHEP12(2015)033}{\emph{JHEP} {\bfseries 12} (2015) 033} [\href{https://arxiv.org/abs/1509.06918}{{\ttfamily 1509.06918}}].

\bibitem{Cicoli:2015ylx}
M.~Cicoli, F.~Quevedo and R.~Valandro, \emph{{De Sitter from T-branes}}, \href{https://doi.org/10.1007/JHEP03(2016)141}{\emph{JHEP} {\bfseries 03} (2016) 141} [\href{https://arxiv.org/abs/1512.04558}{{\ttfamily 1512.04558}}].

\bibitem{Cicoli:2012vw}
M.~Cicoli, S.~Krippendorf, C.~Mayrhofer, F.~Quevedo and R.~Valandro, \emph{{D-Branes at del Pezzo Singularities: Global Embedding and Moduli Stabilisation}}, \href{https://doi.org/10.1007/JHEP09(2012)019}{\emph{JHEP} {\bfseries 09} (2012) 019} [\href{https://arxiv.org/abs/1206.5237}{{\ttfamily 1206.5237}}].

\bibitem{Cicoli:2013mpa}
M.~Cicoli, S.~Krippendorf, C.~Mayrhofer, F.~Quevedo and R.~Valandro, \emph{{D3/D7 Branes at Singularities: Constraints from Global Embedding and Moduli Stabilisation}}, \href{https://doi.org/10.1007/JHEP07(2013)150}{\emph{JHEP} {\bfseries 07} (2013) 150} [\href{https://arxiv.org/abs/1304.0022}{{\ttfamily 1304.0022}}].

\bibitem{Cicoli:2013cha}
M.~Cicoli, D.~Klevers, S.~Krippendorf, C.~Mayrhofer, F.~Quevedo and R.~Valandro, \emph{{Explicit de Sitter Flux Vacua for Global String Models with Chiral Matter}}, \href{https://doi.org/10.1007/JHEP05(2014)001}{\emph{JHEP} {\bfseries 05} (2014) 001} [\href{https://arxiv.org/abs/1312.0014}{{\ttfamily 1312.0014}}].

\bibitem{Cicoli:2017shd}
M.~Cicoli, I.n.~Garc\`\i{}a-Etxebarria, C.~Mayrhofer, F.~Quevedo, P.~Shukla and R.~Valandro, \emph{{Global Orientifolded Quivers with Inflation}}, \href{https://doi.org/10.1007/JHEP11(2017)134}{\emph{JHEP} {\bfseries 11} (2017) 134} [\href{https://arxiv.org/abs/1706.06128}{{\ttfamily 1706.06128}}].

\bibitem{Cicoli:2021dhg}
M.~Cicoli, I.n.G.~Etxebarria, F.~Quevedo, A.~Schachner, P.~Shukla and R.~Valandro, \emph{{The Standard Model quiver in de Sitter string compactifications}}, \href{https://doi.org/10.1007/JHEP08(2021)109}{\emph{JHEP} {\bfseries 08} (2021) 109} [\href{https://arxiv.org/abs/2106.11964}{{\ttfamily 2106.11964}}].

\bibitem{Gallego:2017dvd}
D.~Gallego, M.C.D.~Marsh, B.~Vercnocke and T.~Wrase, \emph{{A New Class of de Sitter Vacua in Type IIB Large Volume Compactifications}}, \href{https://doi.org/10.1007/JHEP10(2017)193}{\emph{JHEP} {\bfseries 10} (2017) 193} [\href{https://arxiv.org/abs/1707.01095}{{\ttfamily 1707.01095}}].

\bibitem{Cicoli:2012fh}
M.~Cicoli, A.~Maharana, F.~Quevedo and C.P.~Burgess, \emph{{De Sitter String Vacua from Dilaton-dependent Non-perturbative Effects}}, \href{https://doi.org/10.1007/JHEP06(2012)011}{\emph{JHEP} {\bfseries 06} (2012) 011} [\href{https://arxiv.org/abs/1203.1750}{{\ttfamily 1203.1750}}].

\bibitem{Westphal:2006tn}
A.~Westphal, \emph{{de Sitter string vacua from Kahler uplifting}}, \href{https://doi.org/10.1088/1126-6708/2007/03/102}{\emph{JHEP} {\bfseries 03} (2007) 102} [\href{https://arxiv.org/abs/hep-th/0611332}{{\ttfamily hep-th/0611332}}].

\bibitem{nodS1}
G.~Obied, H.~Ooguri, L.~Spodyneiko and C.~Vafa, \emph{{De Sitter Space and the Swampland}},  \href{https://arxiv.org/abs/1806.08362}{{\ttfamily 1806.08362}}.

\bibitem{nodS2}
S.K.~Garg and C.~Krishnan, \emph{{Bounds on Slow Roll and the de Sitter Swampland}}, \href{https://doi.org/10.1007/JHEP11(2019)075}{\emph{JHEP} {\bfseries 11} (2019) 075} [\href{https://arxiv.org/abs/1807.05193}{{\ttfamily 1807.05193}}].

\bibitem{nodS3}
H.~Ooguri, E.~Palti, G.~Shiu and C.~Vafa, \emph{{Distance and de Sitter Conjectures on the Swampland}}, \href{https://doi.org/10.1016/j.physletb.2018.11.018}{\emph{Phys. Lett. B} {\bfseries 788} (2019) 180} [\href{https://arxiv.org/abs/1810.05506}{{\ttfamily 1810.05506}}].

\bibitem{Maldacena:2000mw}
J.M.~Maldacena and C.~Nunez, \emph{{Supergravity description of field theories on curved manifolds and a no go theorem}}, \href{https://doi.org/10.1142/S0217751X01003937}{\emph{Int. J. Mod. Phys. A} {\bfseries 16} (2001) 822} [\href{https://arxiv.org/abs/hep-th/0007018}{{\ttfamily hep-th/0007018}}].

\bibitem{Green:2011cn}
S.R.~Green, E.J.~Martinec, C.~Quigley and S.~Sethi, \emph{{Constraints on String Cosmology}}, \href{https://doi.org/10.1088/0264-9381/29/7/075006}{\emph{Class. Quant. Grav.} {\bfseries 29} (2012) 075006} [\href{https://arxiv.org/abs/1110.0545}{{\ttfamily 1110.0545}}].

\bibitem{Gautason:2012tb}
F.F.~Gautason, D.~Junghans and M.~Zagermann, \emph{{On Cosmological Constants from alpha'-Corrections}}, \href{https://doi.org/10.1007/JHEP06(2012)029}{\emph{JHEP} {\bfseries 06} (2012) 029} [\href{https://arxiv.org/abs/1204.0807}{{\ttfamily 1204.0807}}].

\bibitem{Kutasov:2015eba}
D.~Kutasov, T.~Maxfield, I.~Melnikov and S.~Sethi, \emph{{Constraining de Sitter Space in String Theory}}, \href{https://doi.org/10.1103/PhysRevLett.115.071305}{\emph{Phys. Rev. Lett.} {\bfseries 115} (2015) 071305} [\href{https://arxiv.org/abs/1504.00056}{{\ttfamily 1504.00056}}].

\bibitem{Quigley:2015jia}
C.~Quigley, \emph{{Gaugino Condensation and the Cosmological Constant}}, \href{https://doi.org/10.1007/JHEP06(2015)104}{\emph{JHEP} {\bfseries 06} (2015) 104} [\href{https://arxiv.org/abs/1504.00652}{{\ttfamily 1504.00652}}].

\bibitem{Dine:2020vmr}
M.~Dine, J.A.P.~Law-Smith, S.~Sun, D.~Wood and Y.~Yu, \emph{{Obstacles to Constructing de Sitter Space in String Theory}}, \href{https://doi.org/10.1007/JHEP02(2021)050}{\emph{JHEP} {\bfseries 02} (2021) 050} [\href{https://arxiv.org/abs/2008.12399}{{\ttfamily 2008.12399}}].

\bibitem{Montero:2020rpl}
M.~Montero, T.~Van~Riet and G.~Venken, \emph{{A dS obstruction and its phenomenological consequences}}, \href{https://doi.org/10.1007/JHEP05(2020)114}{\emph{JHEP} {\bfseries 05} (2020) 114} [\href{https://arxiv.org/abs/2001.11023}{{\ttfamily 2001.11023}}].

\bibitem{Cunillera:2021fbc}
F.~Cunillera, W.T.~Emond, A.~Leh\'ebel and A.~Padilla, \emph{{Quadratic curvature corrections to stringy effective actions and the absence of de Sitter vacua}}, \href{https://doi.org/10.1007/JHEP02(2022)012}{\emph{JHEP} {\bfseries 02} (2022) 012} [\href{https://arxiv.org/abs/2112.05771}{{\ttfamily 2112.05771}}].

\bibitem{vrdS}
U.H.~Danielsson and T.~Van~Riet, \emph{{What if string theory has no de Sitter vacua?}}, \href{https://doi.org/10.1142/S0218271818300070}{\emph{Int. J. Mod. Phys. D} {\bfseries 27} (2018) 1830007} [\href{https://arxiv.org/abs/1804.01120}{{\ttfamily 1804.01120}}].

\bibitem{Cicoli:2018kdo}
M.~Cicoli, S.~De~Alwis, A.~Maharana, F.~Muia and F.~Quevedo, \emph{{De Sitter vs Quintessence in String Theory}}, \href{https://doi.org/10.1002/prop.201800079}{\emph{Fortsch. Phys.} {\bfseries 67} (2019) 1800079} [\href{https://arxiv.org/abs/1808.08967}{{\ttfamily 1808.08967}}].

\bibitem{Cicoli:2021fsd}
M.~Cicoli, F.~Cunillera, A.~Padilla and F.G.~Pedro, \emph{{Quintessence and the Swampland: The Parametrically Controlled Regime of Moduli Space}}, \href{https://doi.org/10.1002/prop.202200009}{\emph{Fortsch. Phys.} {\bfseries 70} (2022) 2200009} [\href{https://arxiv.org/abs/2112.10779}{{\ttfamily 2112.10779}}].

\bibitem{Cicoli:2021skd}
M.~Cicoli, F.~Cunillera, A.~Padilla and F.G.~Pedro, \emph{{Quintessence and the Swampland: The Numerically Controlled Regime of Moduli Space}}, \href{https://doi.org/10.1002/prop.202200008}{\emph{Fortsch. Phys.} {\bfseries 70} (2022) 2200008} [\href{https://arxiv.org/abs/2112.10783}{{\ttfamily 2112.10783}}].

\bibitem{Hertzberg:2007wc}
M.P.~Hertzberg, S.~Kachru, W.~Taylor and M.~Tegmark, \emph{{Inflationary Constraints on Type IIA String Theory}}, \href{https://doi.org/10.1088/1126-6708/2007/12/095}{\emph{JHEP} {\bfseries 12} (2007) 095} [\href{https://arxiv.org/abs/0711.2512}{{\ttfamily 0711.2512}}].

\bibitem{Garg:2018zdg}
S.K.~Garg, C.~Krishnan and M.~Zaid~Zaz, \emph{{Bounds on Slow Roll at the Boundary of the Landscape}}, \href{https://doi.org/10.1007/JHEP03(2019)029}{\emph{JHEP} {\bfseries 03} (2019) 029} [\href{https://arxiv.org/abs/1810.09406}{{\ttfamily 1810.09406}}].

\bibitem{Ibe:2018ffn}
M.~Ibe, M.~Yamazaki and T.T.~Yanagida, \emph{{Quintessence Axion Revisited in Light of Swampland Conjectures}}, \href{https://doi.org/10.1088/1361-6382/ab5197}{\emph{Class. Quant. Grav.} {\bfseries 36} (2019) 235020} [\href{https://arxiv.org/abs/1811.04664}{{\ttfamily 1811.04664}}].

\bibitem{ValeixoBento:2020ujr}
B.~Valeixo~Bento, D.~Chakraborty, S.L.~Parameswaran and I.~Zavala, \emph{{Dark Energy in String Theory}}, \href{https://doi.org/10.22323/1.376.0123}{\emph{PoS} {\bfseries CORFU2019} (2020) 123} [\href{https://arxiv.org/abs/2005.10168}{{\ttfamily 2005.10168}}].

\bibitem{HebQ}
A.~Hebecker, T.~Skrzypek and M.~Wittner, \emph{{The $F$-term Problem and other Challenges of Stringy Quintessence}}, \href{https://doi.org/10.1007/JHEP11(2019)134}{\emph{JHEP} {\bfseries 11} (2019) 134} [\href{https://arxiv.org/abs/1909.08625}{{\ttfamily 1909.08625}}].

\bibitem{Calderon-Infante:2022nxb}
J.~Calder\'on-Infante, I.~Ruiz and I.~Valenzuela, \emph{{Asymptotic Accelerated Expansion in String Theory and the Swampland}},  \href{https://arxiv.org/abs/2209.11821}{{\ttfamily 2209.11821}}.

\bibitem{Shiu:2023nph}
G.~Shiu, F.~Tonioni and H.V.~Tran, \emph{{Accelerating universe at the end of time}}, \href{https://doi.org/10.1103/PhysRevD.108.063527}{\emph{Phys. Rev. D} {\bfseries 108} (2023) 063527} [\href{https://arxiv.org/abs/2303.03418}{{\ttfamily 2303.03418}}].

\bibitem{Shiu:2023fhb}
G.~Shiu, F.~Tonioni and H.V.~Tran, \emph{{Late-time attractors and cosmic acceleration}}, \href{https://doi.org/10.1103/PhysRevD.108.063528}{\emph{Phys. Rev. D} {\bfseries 108} (2023) 063528} [\href{https://arxiv.org/abs/2306.07327}{{\ttfamily 2306.07327}}].

\bibitem{Cicoli:2020cfj}
M.~Cicoli, G.~Dibitetto and F.G.~Pedro, \emph{{New accelerating solutions in late-time cosmology}}, \href{https://doi.org/10.1103/PhysRevD.101.103524}{\emph{Phys. Rev. D} {\bfseries 101} (2020) 103524} [\href{https://arxiv.org/abs/2002.02695}{{\ttfamily 2002.02695}}].

\bibitem{Cicoli:2020noz}
M.~Cicoli, G.~Dibitetto and F.G.~Pedro, \emph{{Out of the Swampland with Multifield Quintessence?}}, \href{https://doi.org/10.1007/JHEP10(2020)035}{\emph{JHEP} {\bfseries 10} (2020) 035} [\href{https://arxiv.org/abs/2007.11011}{{\ttfamily 2007.11011}}].

\bibitem{Shiu:2024sbe}
G.~Shiu, F.~Tonioni and H.V.~Tran, \emph{{Analytic bounds on late-time axion-scalar cosmologies}},  \href{https://arxiv.org/abs/2406.17030}{{\ttfamily 2406.17030}}.

\bibitem{Brinkmann:2022oxy}
M.~Brinkmann, M.~Cicoli, G.~Dibitetto and F.G.~Pedro, \emph{{Stringy multifield quintessence and the Swampland}}, \href{https://doi.org/10.1007/JHEP11(2022)044}{\emph{JHEP} {\bfseries 11} (2022) 044} [\href{https://arxiv.org/abs/2206.10649}{{\ttfamily 2206.10649}}].

\bibitem{Hebecker:2019csg}
A.~Hebecker, T.~Skrzypek and M.~Wittner, \emph{{The $F$-term Problem and other Challenges of Stringy Quintessence}}, \href{https://doi.org/10.1007/JHEP11(2019)134}{\emph{JHEP} {\bfseries 11} (2019) 134} [\href{https://arxiv.org/abs/1909.08625}{{\ttfamily 1909.08625}}].

\bibitem{Cicoli:2008gp}
M.~Cicoli, C.P.~Burgess and F.~Quevedo, \emph{{Fibre Inflation: Observable Gravity Waves from IIB String Compactifications}}, \href{https://doi.org/10.1088/1475-7516/2009/03/013}{\emph{JCAP} {\bfseries 03} (2009) 013} [\href{https://arxiv.org/abs/0808.0691}{{\ttfamily 0808.0691}}].

\bibitem{Broy:2015zba}
B.J.~Broy, D.~Ciupke, F.G.~Pedro and A.~Westphal, \emph{{Starobinsky-Type Inflation from $\alpha'$-Corrections}}, \href{https://doi.org/10.1088/1475-7516/2016/01/001}{\emph{JCAP} {\bfseries 01} (2016) 001} [\href{https://arxiv.org/abs/1509.00024}{{\ttfamily 1509.00024}}].

\bibitem{Burgess:2016owb}
C.P.~Burgess, M.~Cicoli, S.~de~Alwis and F.~Quevedo, \emph{{Robust Inflation from Fibrous Strings}}, \href{https://doi.org/10.1088/1475-7516/2016/05/032}{\emph{JCAP} {\bfseries 05} (2016) 032} [\href{https://arxiv.org/abs/1603.06789}{{\ttfamily 1603.06789}}].

\bibitem{Cicoli:2016chb}
M.~Cicoli, D.~Ciupke, S.~de~Alwis and F.~Muia, \emph{{$\alpha'$ Inflation: moduli stabilisation and observable tensors from higher derivatives}}, \href{https://doi.org/10.1007/JHEP09(2016)026}{\emph{JHEP} {\bfseries 09} (2016) 026} [\href{https://arxiv.org/abs/1607.01395}{{\ttfamily 1607.01395}}].

\bibitem{Cicoli:2016xae}
M.~Cicoli, F.~Muia and P.~Shukla, \emph{{Global Embedding of Fibre Inflation Models}}, \href{https://doi.org/10.1007/JHEP11(2016)182}{\emph{JHEP} {\bfseries 11} (2016) 182} [\href{https://arxiv.org/abs/1611.04612}{{\ttfamily 1611.04612}}].

\bibitem{Cicoli:2017axo}
M.~Cicoli, D.~Ciupke, V.A.~Diaz, V.~Guidetti, F.~Muia and P.~Shukla, \emph{{Chiral Global Embedding of Fibre Inflation Models}}, \href{https://doi.org/10.1007/JHEP11(2017)207}{\emph{JHEP} {\bfseries 11} (2017) 207} [\href{https://arxiv.org/abs/1709.01518}{{\ttfamily 1709.01518}}].

\bibitem{Bera:2024sbx}
S.~Bera, D.~Chakraborty, G.K.~Leontaris and P.~Shukla, \emph{{Global Embedding of Fibre Inflation in Perturbative LVS}},  \href{https://arxiv.org/abs/2406.01694}{{\ttfamily 2406.01694}}.

\bibitem{Bansal:2024uzr}
S.~Bansal, L.~Brunelli, M.~Cicoli, A.~Hebecker and R.~Kuespert, \emph{{Loop Blow-up Inflation}},  \href{https://arxiv.org/abs/2403.04831}{{\ttfamily 2403.04831}}.

\bibitem{KL}
R.~Kallosh and A.D.~Linde, \emph{{Landscape, the scale of SUSY breaking, and inflation}}, \href{https://doi.org/10.1088/1126-6708/2004/12/004}{\emph{JHEP} {\bfseries 12} (2004) 004} [\href{https://arxiv.org/abs/hep-th/0411011}{{\ttfamily hep-th/0411011}}].

\bibitem{Linde:2007jn}
A.D.~Linde and A.~Westphal, \emph{{Accidental Inflation in String Theory}}, \href{https://doi.org/10.1088/1475-7516/2008/03/005}{\emph{JCAP} {\bfseries 03} (2008) 005} [\href{https://arxiv.org/abs/0712.1610}{{\ttfamily 0712.1610}}].

\bibitem{Conlon:2008cj}
J.P.~Conlon, R.~Kallosh, A.D.~Linde and F.~Quevedo, \emph{{Volume Modulus Inflation and the Gravitino Mass Problem}}, \href{https://doi.org/10.1088/1475-7516/2008/09/011}{\emph{JCAP} {\bfseries 09} (2008) 011} [\href{https://arxiv.org/abs/0806.0809}{{\ttfamily 0806.0809}}].

\bibitem{Cicoli:2015wja}
M.~Cicoli, F.~Muia and F.G.~Pedro, \emph{{Microscopic Origin of Volume Modulus Inflation}}, \href{https://doi.org/10.1088/1475-7516/2015/12/040}{\emph{JCAP} {\bfseries 12} (2015) 040} [\href{https://arxiv.org/abs/1509.07748}{{\ttfamily 1509.07748}}].

\bibitem{Antoniadis:2020stf}
I.~Antoniadis, O.~Lacombe and G.K.~Leontaris, \emph{{Inflation near a metastable de Sitter vacuum from moduli stabilisation}}, \href{https://doi.org/10.1140/epjc/s10052-020-08581-9}{\emph{Eur. Phys. J. C} {\bfseries 80} (2020) 1014} [\href{https://arxiv.org/abs/2007.10362}{{\ttfamily 2007.10362}}].

\bibitem{Antoniadis:2021lhi}
I.~Antoniadis, O.~Lacombe and G.K.~Leontaris, \emph{{Hybrid inflation and waterfall field in string theory from D7-branes}}, \href{https://doi.org/10.1007/JHEP01(2022)011}{\emph{JHEP} {\bfseries 01} (2022) 011} [\href{https://arxiv.org/abs/2109.03243}{{\ttfamily 2109.03243}}].

\bibitem{Bera:2024zsk}
S.~Bera, D.~Chakraborty, G.K.~Leontaris and P.~Shukla, \emph{{Inflating in perturbative LVS: Global Embedding and Robustness}},  \href{https://arxiv.org/abs/2405.06738}{{\ttfamily 2405.06738}}.

\bibitem{DESI:2024mwx}
{\scshape DESI} collaboration, \emph{{DESI 2024 VI: Cosmological Constraints from the Measurements of Baryon Acoustic Oscillations}},  \href{https://arxiv.org/abs/2404.03002}{{\ttfamily 2404.03002}}.

\bibitem{Balasubramanian:2005zx}
V.~Balasubramanian, P.~Berglund, J.P.~Conlon and F.~Quevedo, \emph{{Systematics of moduli stabilisation in Calabi-Yau flux compactifications}}, \href{https://doi.org/10.1088/1126-6708/2005/03/007}{\emph{JHEP} {\bfseries 03} (2005) 007} [\href{https://arxiv.org/abs/hep-th/0502058}{{\ttfamily hep-th/0502058}}].

\bibitem{Cicoli:2008va}
M.~Cicoli, J.P.~Conlon and F.~Quevedo, \emph{{General Analysis of LARGE Volume Scenarios with String Loop Moduli Stabilisation}}, \href{https://doi.org/10.1088/1126-6708/2008/10/105}{\emph{JHEP} {\bfseries 10} (2008) 105} [\href{https://arxiv.org/abs/0805.1029}{{\ttfamily 0805.1029}}].

\bibitem{Cicoli:2011it}
M.~Cicoli, M.~Kreuzer and C.~Mayrhofer, \emph{{Toric K3-Fibred Calabi-Yau Manifolds with del Pezzo Divisors for String Compactifications}}, \href{https://doi.org/10.1007/JHEP02(2012)002}{\emph{JHEP} {\bfseries 02} (2012) 002} [\href{https://arxiv.org/abs/1107.0383}{{\ttfamily 1107.0383}}].

\bibitem{Cicoli:2018cgu}
M.~Cicoli and G.A.~Piovano, \emph{{Reheating and Dark Radiation after Fibre Inflation}}, \href{https://doi.org/10.1088/1475-7516/2019/02/048}{\emph{JCAP} {\bfseries 02} (2019) 048} [\href{https://arxiv.org/abs/1809.01159}{{\ttfamily 1809.01159}}].

\bibitem{Cicoli:2023njy}
M.~Cicoli, M.~Licheri, P.~Piantadosi, F.~Quevedo and P.~Shukla, \emph{{Higher derivative corrections to string inflation}}, \href{https://doi.org/10.1007/JHEP02(2024)115}{\emph{JHEP} {\bfseries 02} (2024) 115} [\href{https://arxiv.org/abs/2309.11697}{{\ttfamily 2309.11697}}].

\bibitem{Cicoli:2022uqa}
M.~Cicoli, K.~Sinha and R.~Wiley~Deal, \emph{{The dark universe after reheating in string inflation}}, \href{https://doi.org/10.1007/JHEP12(2022)068}{\emph{JHEP} {\bfseries 12} (2022) 068} [\href{https://arxiv.org/abs/2208.01017}{{\ttfamily 2208.01017}}].

\bibitem{Cicoli:2018asa}
M.~Cicoli, V.A.~Diaz and F.G.~Pedro, \emph{{Primordial Black Holes from String Inflation}}, \href{https://doi.org/10.1088/1475-7516/2018/06/034}{\emph{JCAP} {\bfseries 06} (2018) 034} [\href{https://arxiv.org/abs/1803.02837}{{\ttfamily 1803.02837}}].

\bibitem{Cicoli:2022sih}
M.~Cicoli, F.G.~Pedro and N.~Pedron, \emph{{Secondary GWs and PBHs in string inflation: formation and detectability}}, \href{https://doi.org/10.1088/1475-7516/2022/08/030}{\emph{JCAP} {\bfseries 08} (2022) 030} [\href{https://arxiv.org/abs/2203.00021}{{\ttfamily 2203.00021}}].

\bibitem{Kaloper:2005aj}
N.~Kaloper and L.~Sorbo, \emph{{Of pngb quintessence}}, \href{https://doi.org/10.1088/1475-7516/2006/04/007}{\emph{JCAP} {\bfseries 04} (2006) 007} [\href{https://arxiv.org/abs/astro-ph/0511543}{{\ttfamily astro-ph/0511543}}].

\bibitem{Arkani-Hamed:2006emk}
N.~Arkani-Hamed, L.~Motl, A.~Nicolis and C.~Vafa, \emph{{The String landscape, black holes and gravity as the weakest force}}, \href{https://doi.org/10.1088/1126-6708/2007/06/060}{\emph{JHEP} {\bfseries 06} (2007) 060} [\href{https://arxiv.org/abs/hep-th/0601001}{{\ttfamily hep-th/0601001}}].

\bibitem{Blumenhagen:2008ji}
R.~Blumenhagen and M.~Schmidt-Sommerfeld, \emph{{Power Towers of String Instantons for N=1 Vacua}}, \href{https://doi.org/10.1088/1126-6708/2008/07/027}{\emph{JHEP} {\bfseries 07} (2008) 027} [\href{https://arxiv.org/abs/0803.1562}{{\ttfamily 0803.1562}}].

\bibitem{Blumenhagen:2012kz}
R.~Blumenhagen, X.~Gao, T.~Rahn and P.~Shukla, \emph{{A Note on Poly-Instanton Effects in Type IIB Orientifolds on Calabi-Yau Threefolds}}, \href{https://doi.org/10.1007/JHEP06(2012)162}{\emph{JHEP} {\bfseries 06} (2012) 162} [\href{https://arxiv.org/abs/1205.2485}{{\ttfamily 1205.2485}}].

\bibitem{Blumenhagen:2012ue}
R.~Blumenhagen, X.~Gao, T.~Rahn and P.~Shukla, \emph{{Moduli Stabilization and Inflationary Cosmology with Poly-Instantons in Type IIB Orientifolds}}, \href{https://doi.org/10.1007/JHEP11(2012)101}{\emph{JHEP} {\bfseries 11} (2012) 101} [\href{https://arxiv.org/abs/1208.1160}{{\ttfamily 1208.1160}}].

\bibitem{Lust:2013kt}
D.~L\"ust and X.~Zhang, \emph{{Four Kahler Moduli Stabilisation in type IIB Orientifolds with K3-fibred Calabi-Yau threefold compactification}}, \href{https://doi.org/10.1007/JHEP05(2013)051}{\emph{JHEP} {\bfseries 05} (2013) 051} [\href{https://arxiv.org/abs/1301.7280}{{\ttfamily 1301.7280}}].

\bibitem{Cicoli:2012tz}
M.~Cicoli, F.G.~Pedro and G.~Tasinato, \emph{{Natural Quintessence in String Theory}}, \href{https://doi.org/10.1088/1475-7516/2012/07/044}{\emph{JCAP} {\bfseries 07} (2012) 044} [\href{https://arxiv.org/abs/1203.6655}{{\ttfamily 1203.6655}}].

\bibitem{hep-ph/0409138}
J.E.~Kim, H.P.~Nilles and M.~Peloso, \emph{{Completing natural inflation}}, \href{https://doi.org/10.1088/1475-7516/2005/01/005}{\emph{JCAP} {\bfseries 01} (2005) 005} [\href{https://arxiv.org/abs/hep-ph/0409138}{{\ttfamily hep-ph/0409138}}].

\bibitem{Long:2014dta}
C.~Long, L.~McAllister and P.~McGuirk, \emph{{Aligned Natural Inflation in String Theory}}, \href{https://doi.org/10.1103/PhysRevD.90.023501}{\emph{Phys. Rev. D} {\bfseries 90} (2014) 023501} [\href{https://arxiv.org/abs/1404.7852}{{\ttfamily 1404.7852}}].

\bibitem{2106.09853}
S.~Angus, K.-S.~Choi and C.S.~Shin, \emph{{Aligned natural inflation in the Large Volume Scenario}}, \href{https://doi.org/10.1007/JHEP10(2021)248}{\emph{JHEP} {\bfseries 10} (2021) 248} [\href{https://arxiv.org/abs/2106.09853}{{\ttfamily 2106.09853}}].

\bibitem{Rudelius:2015xta}
T.~Rudelius, \emph{{Constraints on Axion Inflation from the Weak Gravity Conjecture}}, \href{https://doi.org/10.1088/1475-7516/2015/9/020}{\emph{JCAP} {\bfseries 09} (2015) 020} [\href{https://arxiv.org/abs/1503.00795}{{\ttfamily 1503.00795}}].

\bibitem{oguiso}
K.~Oguiso, \emph{{On Algebraic Fiber Space Structures on a Calabi-Yau 3-fold}}, \href{https://doi.org/10.1142/S0129167X93000248}{\emph{International Journal of Mathematics} {\bfseries 04} (1993) 493}.

\bibitem{Becker:2002nn}
K.~Becker, M.~Becker, M.~Haack and J.~Louis, \emph{{Supersymmetry breaking and alpha-prime corrections to flux induced potentials}}, \href{https://doi.org/10.1088/1126-6708/2002/06/060}{\emph{JHEP} {\bfseries 06} (2002) 060} [\href{https://arxiv.org/abs/hep-th/0204254}{{\ttfamily hep-th/0204254}}].

\bibitem{Minasian:2015bxa}
R.~Minasian, T.G.~Pugh and R.~Savelli, \emph{{F-theory at order $\alpha'^3$}}, \href{https://doi.org/10.1007/JHEP10(2015)050}{\emph{JHEP} {\bfseries 10} (2015) 050} [\href{https://arxiv.org/abs/1506.06756}{{\ttfamily 1506.06756}}].

\bibitem{Berg:2005ja}
M.~Berg, M.~Haack and B.~Kors, \emph{{String loop corrections to Kahler potentials in orientifolds}}, \href{https://doi.org/10.1088/1126-6708/2005/11/030}{\emph{JHEP} {\bfseries 11} (2005) 030} [\href{https://arxiv.org/abs/hep-th/0508043}{{\ttfamily hep-th/0508043}}].

\bibitem{vonGersdorff:2005bf}
G.~von Gersdorff and A.~Hebecker, \emph{{Kahler corrections for the volume modulus of flux compactifications}}, \href{https://doi.org/10.1016/j.physletb.2005.08.024}{\emph{Phys. Lett. B} {\bfseries 624} (2005) 270} [\href{https://arxiv.org/abs/hep-th/0507131}{{\ttfamily hep-th/0507131}}].

\bibitem{Berg:2007wt}
M.~Berg, M.~Haack and E.~Pajer, \emph{{Jumping Through Loops: On Soft Terms from Large Volume Compactifications}}, \href{https://doi.org/10.1088/1126-6708/2007/09/031}{\emph{JHEP} {\bfseries 09} (2007) 031} [\href{https://arxiv.org/abs/0704.0737}{{\ttfamily 0704.0737}}].

\bibitem{Cicoli:2007xp}
M.~Cicoli, J.P.~Conlon and F.~Quevedo, \emph{{Systematics of String Loop Corrections in Type IIB Calabi-Yau Flux Compactifications}}, \href{https://doi.org/10.1088/1126-6708/2008/01/052}{\emph{JHEP} {\bfseries 01} (2008) 052} [\href{https://arxiv.org/abs/0708.1873}{{\ttfamily 0708.1873}}].

\bibitem{Gao:2022uop}
X.~Gao, A.~Hebecker, S.~Schreyer and G.~Venken, \emph{{Loops, local corrections and warping in the LVS and other type IIB models}}, \href{https://doi.org/10.1007/JHEP09(2022)091}{\emph{JHEP} {\bfseries 09} (2022) 091} [\href{https://arxiv.org/abs/2204.06009}{{\ttfamily 2204.06009}}].

\bibitem{Ciupke:2015msa}
D.~Ciupke, J.~Louis and A.~Westphal, \emph{{Higher-Derivative Supergravity and Moduli Stabilization}}, \href{https://doi.org/10.1007/JHEP10(2015)094}{\emph{JHEP} {\bfseries 10} (2015) 094} [\href{https://arxiv.org/abs/1505.03092}{{\ttfamily 1505.03092}}].

\bibitem{Blumenhagen:2009qh}
R.~Blumenhagen, M.~Cvetic, S.~Kachru and T.~Weigand, \emph{{D-Brane Instantons in Type II Orientifolds}}, \href{https://doi.org/10.1146/annurev.nucl.010909.083113}{\emph{Ann. Rev. Nucl. Part. Sci.} {\bfseries 59} (2009) 269} [\href{https://arxiv.org/abs/0902.3251}{{\ttfamily 0902.3251}}].

\bibitem{Cicoli:2020bao}
M.~Cicoli and E.~Di~Valentino, \emph{{Fitting string inflation to real cosmological data: The fiber inflation case}}, \href{https://doi.org/10.1103/PhysRevD.102.043521}{\emph{Phys. Rev. D} {\bfseries 102} (2020) 043521} [\href{https://arxiv.org/abs/2004.01210}{{\ttfamily 2004.01210}}].

\bibitem{Planck:2018jri}
{\scshape Planck} collaboration, \emph{{Planck 2018 results. X. Constraints on inflation}}, \href{https://doi.org/10.1051/0004-6361/201833887}{\emph{Astron. Astrophys.} {\bfseries 641} (2020) A10} [\href{https://arxiv.org/abs/1807.06211}{{\ttfamily 1807.06211}}].

\bibitem{Allahverdi:2023nov}
R.~Allahverdi, C.~Arina, M.~Chianese, M.~Cicoli, F.~Maltoni, D.~Massaro et~al., \emph{{Phenomenology of superheavy decaying dark matter from string theory}}, \href{https://doi.org/10.1007/JHEP02(2024)192}{\emph{JHEP} {\bfseries 02} (2024) 192} [\href{https://arxiv.org/abs/2312.00136}{{\ttfamily 2312.00136}}].

\bibitem{Ferrer:1997yz}
F.~Ferrer and J.A.~Grifols, \emph{{Constraints on the mass of the superlight gravitino from the muon anomaly}}, \href{https://doi.org/10.1103/PhysRevD.56.7466}{\emph{Phys. Rev. D} {\bfseries 56} (1997) 7466} [\href{https://arxiv.org/abs/hep-ph/9704444}{{\ttfamily hep-ph/9704444}}].

\bibitem{Aparicio:2014wxa}
L.~Aparicio, M.~Cicoli, S.~Krippendorf, A.~Maharana, F.~Muia and F.~Quevedo, \emph{{Sequestered de Sitter String Scenarios: Soft-terms}}, \href{https://doi.org/10.1007/JHEP11(2014)071}{\emph{JHEP} {\bfseries 11} (2014) 071} [\href{https://arxiv.org/abs/1409.1931}{{\ttfamily 1409.1931}}].

\bibitem{Marsh:2015xka}
D.J.E.~Marsh, \emph{{Axion Cosmology}}, \href{https://doi.org/10.1016/j.physrep.2016.06.005}{\emph{Phys. Rept.} {\bfseries 643} (2016) 1} [\href{https://arxiv.org/abs/1510.07633}{{\ttfamily 1510.07633}}].

\bibitem{Hardwick:2017fjo}
R.J.~Hardwick, V.~Vennin, C.T.~Byrnes, J.~Torrado and D.~Wands, \emph{{The stochastic spectator}}, \href{https://doi.org/10.1088/1475-7516/2017/10/018}{\emph{JCAP} {\bfseries 10} (2017) 018} [\href{https://arxiv.org/abs/1701.06473}{{\ttfamily 1701.06473}}].

\bibitem{AbdusSalam:2020ywo}
S.~AbdusSalam, S.~Abel, M.~Cicoli, F.~Quevedo and P.~Shukla, \emph{{A systematic approach to K\"ahler moduli stabilisation}}, \href{https://doi.org/10.1007/JHEP08(2020)047}{\emph{JHEP} {\bfseries 08} (2020) 047} [\href{https://arxiv.org/abs/2005.11329}{{\ttfamily 2005.11329}}].

\bibitem{Marsh:2021jmi}
D.J.E.~Marsh and S.~Hoof, \emph{{Astrophysical Searches and Constraints}},  \href{https://arxiv.org/abs/2106.08797}{{\ttfamily 2106.08797}}.

\bibitem{Cheung:2014vva}
C.~Cheung and G.N.~Remmen, \emph{{Naturalness and the Weak Gravity Conjecture}}, \href{https://doi.org/10.1103/PhysRevLett.113.051601}{\emph{Phys. Rev. Lett.} {\bfseries 113} (2014) 051601} [\href{https://arxiv.org/abs/1402.2287}{{\ttfamily 1402.2287}}].

\bibitem{Harlow:2022gzl}
D.~Harlow, B.~Heidenreich, M.~Reece and T.~Rudelius, \emph{{The Weak Gravity Conjecture: A Review}},  \href{https://arxiv.org/abs/2201.08380}{{\ttfamily 2201.08380}}.

\end{thebibliography}\endgroup
\end{document}